# Scaling in four dimensional quantum gravity

*J. Ambjørn* and *J. Jurkiewicz*[1]

The Niels Bohr Institute
Blegdamsvej 17, DK-2100 Copenhagen Ø, Denmark

## Abstract

We discuss scaling relations in four dimensional simplicial quantum gravity. Using numerical results obtained with a new algorithm called "baby universe surgery" we study the critical region of the theory. The position of the phase transition is given with high accuracy and some critical exponents are measured. Their values prove that the transition is continuous. We discuss the properties of two distinct phases of the theory. For large values of the bare gravitational coupling constant the internal Hausdorff dimension is *two* (the elongated phase), and the continuum theory is that of so called branched polymers. For small values of the bare gravitational coupling constant the internal Hausdorff dimension seems to be *infinite* (the crumpled phase). We conjecture that this phase corresponds to a theory of topological gravity. *At* the transition point the Hausdorff dimension might be finite and larger than two. This transition point is a potential candidate for a non-perturbative theory of quantum gravity.

---

[1]Permanent address: Institute of Physics, Jagellonian University, Reymonta 4, 30-059 Krakow, Poland.

hep-th/9503006 01 Mar 1995

# 1 Introduction

Many aspects of two-dimensional quantum gravity are by now well understood. Using Liouville theory [1] we can understand how to quantize two-dimensional quantum gravity coupled to conformal field theories with a central charge $c < 1$. In addition it is understood how to discretize the theory in a way consistent with the underlying reparametrization invariance of the continuum theory [2–5]. The continuum theory is recovered at the critical point of discretized theory. In this way it was possible to utilize the connection between critical phenomena in solid state physics and quantum field theory, which previously has proven so useful. The unique situation if $c < 1$ is that a number of the discretized theories allow an explicit solution. In this way it is possible to compare in detail the exact results with the results of numerical simulations of the discretized systems. Monte Carlo simulations have always been an important tool in the theory of critical phenomena and it can be verified that such methods work very well for two-dimensional quantum gravity coupled to conformal matter too (for a recent review, see [6]). This observation allows us to use numerical simulations in situations where two-dimensional gravity coupled to matter cannot be solved analytically. The numerical approach provided us with a good understanding of the conformal field theories with $c \gg 1$ coupled to quantum gravity [7], and it served as an important inspiration for analytic approximations in this region [8–11].

The discretized version of two-dimensional quantum gravity mentioned above is known as the theory of dynamical triangulations or simplicial quantum gravity. It can be generalized to higher dimensions (in the three-dimensional case see [12,13], in the four-dimensional case [14–24]). By such a generalization we enter unchartered territory, since we have no continuum theory of Euclidean quantum gravity with which we can compare. Nevertheless the generalization is so simple and aesthetically appealing that it deserves a serious investigation. It has not yet been possible to use the strong analytic methods available in two dimensions but as we show in this article, many aspects of the theory can be investigated to a sufficient degree of precision by computer simulations.

The rest or the article is organized as follows: In section 2 we review the definition of the model. In section 3 we discuss the general scaling relations which should be fulfilled if a continuum limit exists in the conventional setting of critical phenomena. These scaling relations are necessary in order to understand the critical behavior of the model. In section 4 we report on the extensive numerical results obtained so far. Subsection 4.1 is a description of a new algorithm for Monte Carlo simulation. It is very efficient in so-called elongated phase where it eliminates the critical slowing down and allows us to determine critical exponents with great precision. Finally section 5 contains a discussion of results and the future perspectives.

# 2 The model

The idea behind simplicial quantum gravity in more than two dimensions goes back to Weingarten [25] and found its "modern" form in [12] in three dimensions and in [14,15] in four dimensions [2]. In the continuum formulation the task is to perform the path integral

---

[2]See [16, 18] for reviews.

over equivalence classes of metrics with the action given by that of general relativity:

$$Z(\Lambda, G, \Sigma, h) = \sum_{\mathcal{M}(\Sigma) \in \text{Top}} \int_{\mathcal{M}(\Sigma)} \frac{\mathcal{D}g_{\mu\nu}}{\text{Vol(diff)}} \, e^{-S_g[\Lambda, G, \Sigma]}, \tag{1}$$

$$S_g[G, \Lambda, \Sigma] = \int_{\mathcal{M}(\Sigma)} d^4\xi \sqrt{g} \left( \Lambda - \frac{1}{16\pi G} R \right) + \text{boundary terms.} \tag{2}$$

Here $\Sigma$ denotes a three-dimensional boundary and the boundary terms depend on the induced metric $h_{\mu\nu}$ on the boundary and the extrinsic curvature induced on $\Sigma$. It is needed in order that the action satisfies the equation:

$$S_{g_1} + S_{g_2} = S_{g_1 + g_2} \tag{3}$$

which is a shorthand notation for the following: The sum of the action for a manifold with metric $g_1$ and boundary $\Sigma$ and induced metric $h$ and the action of a different manifold with metric $g_2$ and opposite oriented boundary $-\Sigma$ and induced metric $h$ must be equal to the action of the manifold constructed by gluing together the two manifolds along $\Sigma$. The condition (3) is needed in order that the partition function $Z(\Sigma_1, h_1; \Sigma_2, h_2)$ satisfies the fundamental composition law:

$$Z(\Sigma_1, h_1; \Sigma_3, h_3) = \sum_{\Sigma_2, h_2} Z(\Sigma_1, h_1; \Sigma_2, h_2) \, Z(\Sigma_2, h_2; \Sigma_3, h_3), \tag{4}$$

which expresses the fact that we should be able to sum over intermediate states, characterized here by the boundary $\Sigma_2$ and the equivalence class of three-metrics $h_2(\Sigma_2)$.

In (2) we are instructed to perform the summation over all manifolds with fixed boundaries and for each manifold to perform the integration over equivalence classes of Riemannian metrics. Nobody has yet been able to make sense of this expression. There is a number of reasons for this. Firstly it is not clear what is meant by the summation $\sum_{\text{Top}}$ over topologies. In four dimensions there is no equivalence between smooth structures and topological structures. There even exist compact topological manifolds with a countable infinity of non-equivalent smooth structures and there exist topological manifolds which do not admit any smooth structure. If we restrict ourselves to smooth structures we have the additional problem that the Euclidean action is unbounded from below. This implies that it is not even clear how to define the path integral over equivalence classes of metrics in the case of a fixed structure. Even if we could ignore the problem of the unboundedness of the Euclidean Einstein action we would still be faced with the problem of evaluation of the path integral. Apart from the technical difficulties of dealing with reparametrization invariant theories we also face the unpleasant fact that the Einstein action is not renormalizable.

Clearly these problems await the proper non-perturbative formulation, either by being embedded in a larger theory (like string theory), or by some discretization method which can tame (1). Simplicial gravity is a suggestion of a discretization which allows us to discuss some aspects of (1) in an non-perturbative setting. It attempts to approximate the integration over equivalence classes by the summation over combinatorially equivalent piecewise linear manifolds.

Here the following should be noted: in four dimensions the equivalence classes of piecewise linear structures are in one-to-one correspondence with smooth structures. If by $\sum_{\text{Top}}$ we understand the summation over all smooth structures, we change nothing by replacing "smooth" with "piecewise linear". To each abstract piecewise linear structure we can associate a metric in the following way: Consider each link in the triangulation to be of equal length $a$ (the "lattice spacing"). If we consider the interior of each four simplex as flat we have defined the concept of distance on the piecewise linear manifold, and in addition Regge calculus associates in a natural way an action to such a piecewise linear manifold. In this case the expression for the action becomes very simple. The volume is proportional to the number of four-simplexes and the curvature term, which is the sum of deficit angles of the two-simplexes, will be proportional to the number of four-simplexes containing the two-simplex. For a closed manifold the sum over all deficit angles can now be expressed as a linear combination of the total number of four-simplexes, $N_4$, and the total number of two-simplexes, $N_2$. If we take the lattice unit $a = 1$ the continuum action (2) for a piecewise linear manifold characterized by an abstract triangulation $T$ with boundary $B$ takes the exceedingly simple form:

$$S_T[k_4, k_2] = k_4 N_4(T) - k_2 N_2(T) + \frac{1}{2} k_2 N_2(B), \tag{5}$$

where $k_2 \sim 1/G$. One could ask if such a simple expression has any chance of representing a theory of quantum gravity. Our attitude is the opposite: It is intriguing that there exists such a simple expression and hopefully the simplicity might lead to the explicit solution of quantum gravity by analytic means in the future. This is precisely what has happened in two dimensional quantum gravity.

The path integral (1) is replaced by

$$Z[k_2, k_4] = \sum_T \frac{1}{C_T} e^{-S_T[k_2, k_4]}, \tag{6}$$

where $C_T$ is a symmetry factor of the graph $T$ (the order of the automorphism group of the graph). The integration over equivalence classes of metrics in this formula has been replaced by the summation over a grid in the space of equivalence classes of metrics. Each abstract triangulation (up to graph isomorphism) represents such a point, the metric being assigned as described above. Different triangulations (up to graph isomorphism) correspond to inequivalent metrics and consequently to different points. The weight of the given metric (or triangulation) is given by the exponential of the action (the Boltzmann weight), calculated by Regge's prescription. The basic assumption is that this grid (with Boltzmann weight included) becomes dense in the total space of equivalence classes of metrics (with Boltzmann weight included) for certain values of the bare coupling constants $k_2, k_4$, such that the sum (6) provides us with a realization of (1).

One annoying aspect of (1) is the split in a summation over topologies and the integration over Riemannian structures. It seems that in a natural way we have been able to combine these in simplicial quantum gravity by formula (6). This is yet another feature of simplicial quantum gravity which has been very important in the two-dimensional models, where the summation over topologies, known as the double scaling limit, can be performed in the simplicial quantum gravity models. Unfortunately the sum (6) is still quite formal, and in more that two dimensions it is not yet known how to make sense of it. If we insist on an unrestricted sum over topologies there is a huge entropy factor, which is not sufficiently damped by the action, no matter what are the values of the bare coupling constants $k_2, k_4$. In this paper we shall make no attempt to define the sum over topologies, but confine ourself to the simplest of all cases, that of four-manifolds of spherical topology, except for simple boundaries each of a topology corresponding to $S^3$.

At the moment we restrict the topology in the summation (6) the sum becomes well defined in some range of the $(k_2, k_4)$ coupling constant plane. This follows from the fact that the number of triangulations of a given topology seems to be exponentially bounded with the volume, i.e. the number of four-simplexes. The analytic proof is well known in two dimensions [26]. In higher dimensions recently there has been some controversy about this point [27]. Computer simulations support the exponential bound [28,21] and very recently there appeared an analytic proof of the existence of the exponential bound in all dimensions and for all (fixed) topologies [31]. The existence of the exponential bound implies that the number of triangulations $\mathcal{N}(N_4)$ for the number of four-simplexes equal to $N_4$ (and for fixed topology) will be bounded by

$$\mathcal{N}(N_4) \leq e^{cN_4}. \tag{7}$$

Since we have a trivial bound $N_2(T) < \text{constant} \cdot N_4(T)$ it follows that for any fixed $k_2$ (6) is well defined for sufficiently large $k_4$. For a fixed $k_2$ there is a critical lower value $k_4^c(k_2)$ of $k_4$ such that the partition function (6) is well defined for $k_4 > k_4^c(k_2)$ and divergent for $k_4 < k_4^c(k_2)$. To be more specific we can introduce the partition function for fixed (lattice) volume:

$$Z(k_2, N_4) = \sum_{T \in \mathcal{T}(N_4)} e^{k_2 N_2(T)} = f_{k_2}(N_4) e^{k_4^c(k_2)N_4}, \tag{8}$$

where $\mathcal{T}(N_4)$ denotes the class of triangulations of fixed topology and boundary conditions, here always taken to be that of a four-sphere with $S^3$ boundaries. The function $f_{k_2}(N_4)$ denotes a subleading factor and it will be discussed in detail later. Here it is sufficient to say that it will depend on the boundary conditions and on $k_2$.

In the limit $k_4 \to k_4^c(k_2)$ large $N_4$ will dominate (6). *It corresponds to the infinite (lattice) volume limit.* This implies that the infinite volume limit is found by the approach to a the critical line $k_4 = k_4^c(k_2)$ in the $(k_2, k_4)$ coupling constant plane. It should be emphasized that the existence of an infinite volume limit does not necessarily imply that we have an interesting continuum limit. Rather, the situation should be compared with a lattice theory. For a spin system we can always take the infinite volume limit by taking the lattice volume to infinity. However, only for a specific range of the coupling constants we shall be able to obtain a spin-spin correlation length which diverges with respect to the lattice spacing, and it is only for this range of coupling constants that we have any hope of obtaining a conventional continuum field theory.

## 3 Scaling relations

In the case of a spin system on a regular lattice the spin-spin correlation length, or the mass gap $m(\beta)$, is the quantity which characterizes the scaling behavior. If $m(\beta) \to 0$ for $\beta \to \beta_c$, where $\beta_c$ denotes the critical point, it might be possible to define a continuum limit of the spin system, since the correlation length becomes infinite compared to the lattice spacing. All scaling relations can be derived from this scaling hypothesis. We can introduce a similar object in simplicial quantum gravity.

Let us first discuss the formal continuum physics of the object. In order to address the question of divergent correlation length we have first to introduce the concept of length in quantum gravity. It can be done even if we have to integrate over the metrics. The simplest example of a reparametrization invariant correlation function which depends on a geodesic distance is the volume-volume correlator:

$$G(r; \Lambda, G) = \left\langle \int \int \sqrt{g(\xi)} d^4\xi \sqrt{g(\xi')} d^4\xi' \, \delta(d_g(\xi, \xi') - r) \right\rangle, \qquad (9)$$

where $d_g(\xi, \xi)$ denote the geodesic distance between the points $\xi$ and $\xi'$, calculated with the metric $g_{\mu\nu}$, and where the average is performed with the formal path integral of quantum gravity, i.e. over all equivalence classes of metrics $[g]$. The subscript $c$ in $\langle \cdot \rangle_c$ denotes the connected part of the correlator. In a similar way one can include in the average reparametrization invariant local operators like the scalar curvature $R(x)$ or $R^2_{\mu\nu\kappa,\rho}$:

$$G_R(r; \Lambda, G) = \left\langle \int \int \sqrt{g(\xi)} R(\xi) d^4\xi \sqrt{g(\xi')} R(\xi') d^4\xi' \, \delta(d_g(\xi, \xi') - r) \right\rangle. \qquad (10)$$

The integrals of $G(r, \Lambda, G)$ and $G_R(r, \Lambda, G)$ are directly related to the partition function $Z(\Lambda, G)$ itself by:

$$\int_0^\infty dr \, G(r; \Lambda, G) = \frac{\partial^2 \log Z(\Lambda, G)}{\partial \Lambda^2} \qquad (11)$$

and

$$\int_0^\infty dr \, G_R(r; \Lambda, G) \sim \frac{\partial^2 \log Z(\Lambda, G)}{\partial(1/G)^2} \qquad (12)$$

In principle one could also consider such correlators for a fixed volume $V$ of the universe, e.g.

$$G(r; V, G) = \left\langle \int \int \sqrt{g(\xi)} d^4\xi \sqrt{g(\xi')} d^4\xi' \, \delta(d_g(\xi, \xi') - r) \delta(\int d^4\xi \sqrt{g} - V) \right\rangle. \qquad (13)$$

While not so interesting from a physical point of view the knowledge of the correlators for any volume allows us to reconstruct the former correlators by e.g.:

$$G(r; \Lambda, G) = \int_0^\infty dV e^{-\Lambda V} G(r; V, G), \quad G(r; V, G) = \int_{c-i\infty}^{c+i\infty} \frac{d\Lambda}{2\pi i} e^{\Lambda V} G(r; \Lambda, G). \qquad (14)$$

From a numerical point of view $G(r; V, G)$ are more convenient objects.

The above mentioned correlation functions have a direct and simple translation to the discretized approximation where all quantities are well defined. Consider piecewise linear manifolds with two marked four-simplexes, or equivalently with two boundaries which are the boundaries of two four-simplexes. The geodesic distance between any two points is uniquely defined from the piecewise linear structure and the requirement that the link length is $a = 1$. Let us for simplicity (in particular for the numerical work to be reported below) adopt a simplified notation[3]: We define a path between two *four-simplexes* as a sequence of neighboring four-simplexes connecting the two. The geodesic distance is the shortest such path. This definition can in an obvious way be used to define the geodesic distance between two boundaries. Let us introduce the notation $\partial s_4$ for the boundary of a four-simplex $s_4$. We can now define the 2-point function as the partition function (1):

$$G(r; k_2, k_4) = \sum_{T \in \mathcal{T}_2(r)} e^{-S_T[k_2, k_4]},$$ (15)

where $\mathcal{T}_2(r)$ denotes the class of triangulations of $S^4$ with two boundaries $\partial s_4(1)$ and $\partial s_4(2)$ separated by a geodesic distance $r$. The important property of $G(r; k_2, k_4)$ is its exponential decay with the geodesic distance $r$. It follows from the inequality:

$$G(r_1 + r_2; k_2, k_4) \geq \text{const} \cdot G(r_1; k_2, k_4) \, G(r_2; k_2, k_4).$$ (16)

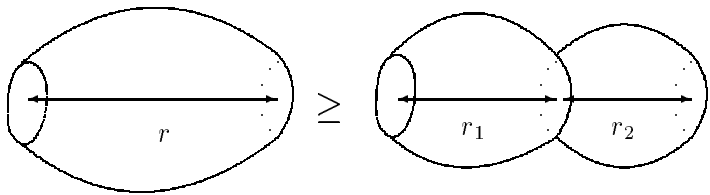

Figure 1: Graphical illustration of inequality (16) in the two dimensional case. Eq. (16) corresponds to the situation where the boundaries are contracted to points, but is in fact valid for arbitrary boundaries.

This inequality has a simple geometrical interpretation. Given two triangulations $T_1$ with boundaries $\partial s_4(1)$ and $\partial s_4(3)$ separated by distance $r_1$, and $T_2$ with boundaries $\partial s_4(4)$ and $\partial s_4(2)$ separated by distance $r_2$, we can clearly construct a triangulation "$T_1 + T_2$" with boundaries $\partial s_4(1)$ and $\partial s_4(2)$ separated by a distance $r_1 + r_2$ by gluing together $T_1$ and $T_2$ along $\partial s_4(3)$ and $\partial s_4(4)$. Due to the property (3) the action for the "$T_1 + T_2$" will be the sum of the actions for $T_1$ and $T_2$. Except for a trivial constant factor of order one due to different ways of gluing $\partial s_4(3)$ and $\partial s_4(4)$ together each $T_1$ and $T_2$ entering on the lhs of eq. (17) is present in the sum on the rhs of (16) in the form "$T_1 + T_2$" with the same Boltzmann weight. In addition there are of course many terms in the sum on the rhs of (16) which have no representation on the lhs of (16). The inequality is shown graphically in fig. 1.

The important implication of (16) is that the function $-\log G(r; k_2, k_4)$ is a subadditive function of $r$. This implies that

$$\lim_{r \to \infty} \frac{-\log G(r; k_2, k_4)}{r} \equiv m(k_2, k_4) \quad \text{exists.}$$ (17)

---

[3]This definition of geodesic distance has the additional advantage that it applies to more general complexes than genuine triangulations. In the broader context of matrix models one will be forced to sum over such general complexes [12] and here it is convenient to be able to use the same concept of geodesic distance.

In addition it is not difficult to show by combinatorial arguments that for $k_4 > k_4^c(k_2)$

$$-\frac{\partial G(r=1;k_2,k_4)}{\partial k_4} \geq c^2 \, G(r;k_2,k_4),$$ (18)

$$-\frac{\partial G(r;k_2,k_4)}{\partial k_4} \geq \tilde{c}^2 \, G(r;k_2,k_4)$$ (19)

where $c^2$ and $\tilde{c}^2$ are two constants of $\mathcal{O}(1)$. The arguments are identical to the ones originally used for random surfaces [32] and they imply that

$$m(k_2,k_4) \geq 0 \quad \text{and} \quad \frac{\partial m(k_2,k_4)}{\partial k_4} > 0 \quad \text{for} \quad k_4 > k_4^c(k_2).$$ (20)

The first inequality follows from (17) and (18) in the limit $r \to \infty$. The second inequality for from (17) and (19) for $r \to \infty$.

*To summarize*: For $k_4 > k_4^c(k_2)$ the 2-point function falls off exponentially, but although the mass $m(k_2,k_4)$ is decreasing for $k_4 \to k_4^c(k_2)$ the general arguments do not tell us that the mass scales to zero.

Let us in the following *assume* that

$$m(k_2,k_4) \sim (k_4 - k_4^c(k_2))^\nu \quad \text{for} \quad k_4 \to k_4^c(k_2),$$ (21)

and deduce the consequences of this assumption. In the following we will suppress the explicit $k_2$ dependence and introduce the shorthand notation $\triangle k_4 \equiv k_4 - k_4^c(k_2)$. Let us further introduce the 2-point function $G(r;N_4)$ for a fixed volume $N_4$:

$$G(r;N_4) = \sum_{T \in \mathcal{T}_2(r;N_4)} e^{k_2 N_2(T)}.$$ (22)

The function is convenient in numerical simulations as will become clear in the following. It is related to $G(r;k_4)$ by a (discrete) Laplace transform:

$$G(r;k_4) = \sum_{N_4} e^{-k_4 N_4} G(r;N_4).$$ (23)

From the large distance behavior of $G(r,k_4)$ dictated by (17) and (21) we deduce the long distance behavior of $G(r;N_4)$ by an inverse Laplace transform of (22):

$$G(r;k_4) \sim e^{-c\,r\,(\triangle k_4)^\nu} \quad \Rightarrow \quad G(r;N_4) \sim e^{-\tilde{c}(r/N_4^\nu)^{1/(1-\nu)}}$$ (24)

*The mass exponent $\nu$ has the interpretation as the inverse internal Hausdorff dimension of the ensemble of four-manifolds*:

$$\nu = 1/d_H.$$ (25)

Let us prove this relation. Given the ensemble of four-manifolds with boundaries $\partial s_4(1)$ and $\partial s_4(2)$ separated by a geodesic distance $r$ it is natural to define the Hausdorff dimension by

$$\langle N_4 \rangle_r \sim r^{d_H}$$ (26)

where $\langle A(N_4)\rangle_r$ is defined by

$$\langle A(N_4)\rangle_r = \frac{\sum_{T\in\mathcal{T}_2(r)} A(N_4(T))\mathrm{e}^{-S_T[k_4]}}{\sum_{T\in\mathcal{T}_2(r)} \mathrm{e}^{-S_T[k_4]}} \qquad (27)$$

For large $r$ we can use the asymptotic form $G(r;k_4) \sim \mathrm{e}^{-m(k_4)r}$ of the 2-point function to calculate the $\langle N_4\rangle$:

$$\langle N_4\rangle_r = -\frac{1}{G(r;k_4)}\frac{dG(r;k_4)}{dk_4} \sim m'(k_4)r. \qquad (28)$$

This formula expresses the trivial fact that for fixed $k_4 > k_4^c$ and $r \to \infty$ the typical manifold is a long tube and the Hausdorff dimension is one. The same would be the case for an ordinary random walk and it highlights the way the scaling limit should be taken: We want $r \to \infty$ in order to be able to ignore lattice artifacts, but at the same time we have to scale $m(k_4) \to 0$ such that the 2-point function $\sim \mathrm{e}^{-m(k_4)r}$ survives, i.e.:

$$m(k_4)r = \mathrm{const.} \quad \mathrm{or} \quad (\triangle k_4)^\nu r = \mathrm{const.} \qquad (29)$$

If we use this in (28) we get:

$$\langle N_4\rangle_r \sim r^{1/\nu}, \quad \mathrm{i.e.} \quad d_H = \frac{1}{\nu}. \qquad (30)$$

One can compare this result with the ordinary random walk in $R^n$ where it is well known that the mass exponent $\nu = 1/2$ and the (*external*) Hausdorff dimension is two.

While $G(r;k_4)$ is a convenient object from a theoretical point of view, it is $G(r;N_4)$ which is readily available from the point of view of computer simulations. Let us assume the Hausdorff dimension of our ensemble of manifolds is $d_H$. An alternative definition of $d_H$ would be the following: Let us consider the ensemble of four-manifolds with a fixed volume $N_4$ and one boundary $\partial s_4$. Let us move away from the boundary in "spherical" shells of thickness 1 and count the number of four-simplexes $n(r)$ in the shells a distance $r$ from the boundary [14,23]. If we take the average over the ensemble of manifolds we expect the following behavior, as long as $r$ is not too large compared to $N_4^{1/d_H}$:

$$\langle n(r)\rangle_{N_4} \sim r^{d_H-1} \quad \mathrm{for} \quad 1 \ll r \ll N_4^{1/d_H}. \qquad (31)$$

The number $n(r)$ is closely related to $G(r;N_4)$. Since we can view $G(r=0;N_4)$ as the partition function for the ensemble of manifolds with one boundary $\partial s_4$ except for a trivial combinatorial factor of order one, we have:

$$\langle n(r)\rangle_{N_4} \sim \frac{G(r;N_4)}{G(0;N_4)}, \qquad (32)$$

from which by (24) we deduce the long distance behavior of $\langle n(r)\rangle_{N_4}$:

$$\langle n(r)\rangle_{N_4} \sim \mathrm{e}^{-\left(rN_4^{1/d_H}\right)^{d_H/(d_H-1)}} \quad \mathrm{for} \quad N_4^{1/d_H} < r \ll N_4. \qquad (33)$$

While the long distance behavior of $G(r; N_4)$ is determined from that of $G(r; k_4)$, which again follows from the scaling assumption (21), the short distance behavior of $G(r; N_4)$ is dictated by (31):

$$G(r; N_4) \sim G(0; N_4) \; r^{d_H - 1} \quad \text{for} \quad 1 \ll r \ll N_4^{1/d_H}, \tag{34}$$

and we can determine the short distance behavior of $G(r; k_4)$ from (23). Recall the expression (8) for the partition function $Z(N_4)$ for finite volume (we still suppress explicit reference to $k_2$). For $N_4$ not too small we have

$$G(0; N_4) \sim N_4 Z(N_4), \tag{35}$$

since $G(0, N_4)$ corresponds to manifolds with one boundary $\partial s_4$ and this boundary can be inserted at any of the $N_4$ simplexes. For very small symmetric triangulations additional combinatorial factors might be present, this is why we say that $N_4$ should not be too small. According to (8) we expect $Z(N_4)$ to grow exponentially with $N_4$. The critical behavior of $G(r, k_4)$ is related to the subleading behavior $f(N_4)$. Two functional forms of the subleading behavior have been suggested:

$$f(N_4) = N_4^{\gamma - 3} \tag{36}$$

and

$$f(N_4) = e^{-c N_4^\alpha}. \tag{37}$$

The critical behavior corresponding to these two possibilities will be very different.

Let us first discuss the implications of the form (36) which is well known from the string theory and two-dimensional simplicial gravity. Like in two dimensions [9, 8] one can prove that either $\gamma \leq 0$ or $\gamma = 1/n$, $n \geq 2$. For $r \ll 1/\triangle k_4$ we get, using (23) and (34)-(36):

$$G(r, k_4) \sim r^{d_H - 1} \sum_{N_4} N_4^{\gamma - 2} \; e^{-\triangle k_4 N_4} \; e^{-\left(r N_4^{1/d_H}\right)^{d_H/(d_H - 1)}} \; \sim \; r^{\gamma d_H - 1}. \tag{38}$$

It follows that the short distance behavior of $G(r, k_4)$ is determined from the Hausdorff dimension and the "entropy exponent"[4] $\gamma$. This behavior reflects both kinds of fractal structures readily definable in quantum gravity: The first is the internal Hausdorff dimension $d_H$ and the second is the baby universe structure.

The fact that the exponent $\gamma$ defined by (36) reflects the distribution of baby universes was shown in [33] in two dimensional simplicial gravity, and generalized to four dimensional simplicial quantum gravity in [19]. Let us first define a minimal bottleneck baby universe, abbreviated "minbu". A "bottleneck" in a triangulation of $S^4$ is a collection of five three-simplexes isomorphic to the boundary $\partial s_4$ of a four-simplex. If we cut our piecewise linear manifold along such a closed three dimensional manifold we will separate it in two. Usually this separation will be trivial in the sense that the three-dimensional manifold will be the boundary of a four-simplex $s_4$ in the triangulation and all we have done is to separate this four-simplex from the rest of the triangulation. However, in case the three-dimensional closed manifold is not the boundary of a four-simplex belonging to the triangulation, cutting along the boundary will produce a separation into two non-trivial parts of the triangulation. The smaller one is called a "minimal bottleneck baby universe", abbreviated "minbu" as mentioned above, the larger the "parent".

---

[4]The notation "entropy exponent" is justified in case where $k_2 = 0$, since $Z(N_4)$ in this case simply counts the number of triangulations.

The average number of minbu's of volume $V_4$ on 4-manifolds of total volume $N_4$ can be expressed in terms of the partition function $Z(N)$ and will be given by a general formula

$$\langle \mathcal{N}(V_4) \rangle_{N_4} \approx \frac{120}{Z(N_4)} V_4 Z(V_4) \ (N_4 - V_4) Z(N_4 - V_4) \tag{39}$$

where 120 is the number of ways one can glue the two boundaries of the minbu and the parent together with opposite orientation of the boundary. The additional factors $V_4$ and $N_4 - V_4$ appear because the minbu and the parent both have a minimal boundary $\partial s_4$, i.e. for a generic large surface of volume $V_4$ and no accidental symmetry factors there will be $V_4$ such manifolds for each closed manifold since the boundary can be placed at any of the $V_4$ four-simplexes. If we *assume* that the canonical partition function is given by (36), we get:

$$\langle \mathcal{N}(V_4) \rangle_{N_4} \sim \ N_4 \ [V_4(1 - V_4/N_4)]^{\gamma - 2} \ . \tag{40}$$

which has the obvious interpretation that the probability of having a branching to a minbu of volume $V_4$ per unit volume is $V_4^{\gamma - 2}(1 - V_4/N_4)^{\gamma - 2}$. As shown in [34,19] "minbu counting" is a very convenient tool for measuring $\gamma$ if $\gamma$ is not too negative, both in two-dimensional and four-dimensional simplicial quantum gravity.

The "entropy exponent" $\gamma$ is also called the "susceptibility exponent". The reason is that it appears in the integral (11) of the two-point function $G(r; k_4)$:

$$\sum_r G(r; k_4) \equiv \chi_V(k_4) = \frac{c}{(k_4 - k_4^c)^\gamma} + \text{less singular terms.} \tag{41}$$

From the continuum formula (11) it follows that

$$\chi_V(k_4) = \frac{d^2 \log Z(k_4)}{dk_4^2} \tag{42}$$

It follows that the volume-volume fluctuations diverge if $\gamma > 0$ and stay finite if $\gamma < 0$.

It is also possible to define the discretized version of the continuum curvature-curvature correlation function (10). The integral of this discretized curvature-curvature correlation function $\chi(k_2)$ has been extensively used in the former studies of possible phase transitions as a function of $k_2$. The analogue to the continuum formula (12) would be:

$$\chi_R(k_2) = \frac{d^2 \log Z(k_2)}{dk_2^2}, \qquad k_4 = k_4^c(k_2), \tag{43}$$

and the existence of a $k_2^c$ where $\chi_R(k_2)$ diverges would be an indication of a phase transition. In this study we will not focus on this quantity, but can still add to the understanding of its the behavior (see section 4.6).

The asymptotic form (37) leads to a scaling behavior different from the one just discussed. Formula (39) is of course valid also in this case. In the following we shall see that there is good evidence that the behavior (37) is related to $d_H = \infty$ (or $\nu \to 0$). From (24) we get:

$$G(r; N_4) \to f(N_4) \; \mathrm{e}^{-\triangle k_4 N_4 - c\, r} \quad \text{for} \quad d_H \to \infty, \quad r > r_0 \tag{44}$$

where $r_0$ is some short distance cut-off of $\mathcal{O}(1)$ with only a weak $N_4$ dependence. The $r$ dependence factorizes and results in the lack of scaling of the mass:

$$G(r; k_4) = \sum_{N_4} G(r; N_4) \; \mathrm{e}^{-k_4 N_4} \sim \mathrm{e}^{-c\, r}. \tag{45}$$

In principle it is still possible to maintain a scaling of the mass $m(k_4)$ for $k_4 \to k_4^c$ even if the Hausdorff dimension is infinite: A behavior like

$$m(k_4) \sim \frac{1}{\log \frac{1}{\triangle k_4}} \tag{46}$$

is possible. However, it will not lead to a factorization of the $r$ dependence in $G(r; N_4)$, but to a dependence:

$$G(r; N_4) \sim \mathrm{e}^{-\frac{r}{\log(N_4/r)}(1 + \mathcal{O}(\frac{\log\log(N_4/r)}{\log(N_4/r)}))}. \tag{47}$$

Although this is a weak $N_4$ dependence we shall be able to distinguish between (45) and (46) in the numerical simulations.

It is clear that it is meaningless to talk about a susceptibility exponent $\gamma$ in case we have the asymptotic behavior (37). Formally $\gamma = -\infty$. While a negative exponent $\gamma$ implies that the susceptibility $\chi_V(k_4)$ of volume fluctuations remains finite at the critical point $k_4^c$, the derivative $\chi_V^n(k_4)$ after $k_4$ will diverge for sufficiently large $n$. This will not be the case if we have the asymptotic behavior (37).

## 4 The computer simulations

Since numerical simulations will be the source of information about the properties of the four-dimensional simplicial systems let us start this section with a discussion of certain theoretical and practical problems connected with Monte Carlo simulations of four dimensional quantum gravity. Given an initial triangulation $T_0(\mathcal{M})$ of a four-manifold $\mathcal{M}$ the Monte Carlo simulations described here make use of two kinds of changes, a standard finite set of so called local moves (there are five different local moves for triangulations of four-manifolds) and global "baby universe surgery" moves described below. The local moves were discussed in a number of articles[5] Each of them can be implemented by a finite number of computer instructions, independently of the size of the triangulation and their weights are chosen as the Boltzmann weights dictated by the classical action of general relativity, i.e. by $\mathrm{e}^{-S_T}$. The local moves are ergodic. By this we mean that given any two triangulations of the manifold it is possible in a finite number of moves, carried out successively, to change one of the triangulations to the other. The local moves seem therefore to provide a perfect setup from a computational point of view.

---

[5]For reviews see e.g. [19, 18].

There is however one serious theoretical problem related to the fact that there exist four-manifolds which are *algorithmically unrecognizable*. Denote such a manifold by $\mathcal{M}_0$ and let it be finitely presented by a triangulation $T(\mathcal{M}_0)$. The algorithmic unrecognizability of $\mathcal{M}_0$ means that there exists no algorithm which allows us to decide whether another manifold $\mathcal{M}$, again finitely presented by a triangulation $T(\mathcal{M})$ is combinatorially equivalent[6] to $\mathcal{M}_0$. When this is combined with the existence of the finite set of local moves which is able to bring us between any two triangulations of $\mathcal{M}_0$ in a finite number of steps, but where this number is a function of the chosen triangulations, one can prove the following theorem [35]:

*The number of moves needed to connect two triangulations of $\mathcal{M}_0$, $T$ and $T'$ with $N_4(T) = N_4(T')$, cannot be bounded by any recursive function $r(N_4)$.*

This theorem implies that there will be very large barriers between some classes of triangulations of $\mathcal{M}_0$ and there would be triangulations which can never be reached in any reasonable number of steps. Of course the number of configurations which are separated from some standard triangulation of $\mathcal{M}_0$ by such barriers could vanish relative to the total number of configurations as a function of $N_4$. In ref. [35] it was conjectured that it will not be the case. If the conjecture is correct a Monte Carlo method based on the finite set of local moves will never get around *effectively* in the class of triangulations of $\mathcal{M}_0$. We can say that the moves, although ergodic in the class of triangulations, will not be *computationally ergodic* [35].

The situation for $S^4$ which is the manifold we use in the actual Monte Carlo simulations was discussed in [28]. It is unknown whether $S^4$ is algorithmically recognizable in the class of four-manifolds. If $S^4$ is algorithmically unrecognizable the arguments given above for $\mathcal{M}_0$ apply. Any attempt to see high barriers separating triangulations for $S^4$ has failed so far[7]. Does this mean that we have numerical evidence that $S^4$ is algorithmically recognizable? Unfortunately not. While the four-dimensional manifolds which are presently known to be algorithmically unrecognizable are rather unwieldy, the situation changes if we move one dimension up, and even a simple manifold like $S^5$ is algorithmically unrecognizable. Computer simulations for $S^5$ give results similar to those of $S^4$, i.e. up to now there has been no trace of any barriers [36]. Two interpretations are possible: Either the class of configurations of $S^5$ which are separated from the rest by high barriers contains very special configurations which are of measure zero and not important for numerical simulations, or the barriers are so high that we have simply not encountered them yet for the size of triangulations which have been used in simulations of $S^5$.

In the following we assume that there are no problems with the Monte Carlo simulations for $S^4$ related to the algorithmic unrecognizability. Since we are interested in studying the scaling relations near the critical point of the theory we are faced with another serious practical problem when using only a standard set of the local moves. As will become clear from the results presented below, the typical geometry of the manifold depends strongly on the value of the gravitational coupling constant $k_2$ and becomes that of a branched polymer for large values of $k_2$. The effect is a critical slowing down which makes the study of the branched polymer phase as well as the critical region very difficult. Local moves need many steps to change the geometric structure of the manifold and we observe very long autocorrelation times for quantities like $\langle n(r) \rangle_{N_4}$ (33). To overcome this difficulty we enlarge the set of moves by global "baby universe surgery" moves which permit large changes of geometry and at the same time have large acceptance (in fact equal one). Similar moves were proposed earlier for two-dimensional systems [37] and led to a large reduction of the autocorrelation times.

---

[6]To be precise the phrase "combinatorial equivalence" means that there exists a triangulation which (up to graph isomorphism) is a common subdivision of the two triangulations.

[7]The way to search for such barriers is the following: Generate by Monte Carlo simulations a number of independent configurations for some large values of $N_4$. Now "shrink" (again by Monte Carlo simulations) these configurations to the minimal triangulation of $S^4$, consisting of 5 4-simplexes. In case we never get seriously stuck in this shrinking procedure there can be no barrier separating two triangulations since we can first move to the minimal configuration and then out to another triangulation by the reverse set of moves. We have never observed that the triangulations get stuck in the process of a "reasonable shrinking procedure".

It should also be noted that the discussion of the algorithmic nonrecognizability deals only with the local moves. The theoretical implications of "baby universe surgery" are not known.

## 4.1 Baby universe surgery

The main new ingredient of our simulations is the "baby universe surgery" algorithm. The basic concept here is that of the "minimal bottleneck baby universe" introduced in Section 3. We recall that a "bottleneck" is a collection of five three-simplexes isomorphic to the boundary of a four-simplex, but not forming a boundary of any four-simplex belonging to the manifold. The "bottleneck" forms a boundary dividing the manifold in two parts. Since the original manifold has a topology of $S^4$ both parts have a structure of $S^4$ with the boundary $\partial s_4$. The manifold can be split into two sub-manifolds with topology of $S^4$ by cutting it along the bottleneck and adding one four-simplex in each part to close the boundary $\partial s_4$. This is the first step of the global "baby universe surgery" move. This step can be inverted: In each sub-manifold we remove a randomly chosen four-simplex, changing the topology of the sub-manifold to be that of $S^4$ with a boundary $\partial s_4$. The two parts can now be glued together and the boundary $\partial s_4$ becomes a new "bottleneck". The gluing can be performed in 5! ways.

One can easily convince oneself that the process described above is self-dual: the numbers of simplexes and sub-simplexes remain unchanged. This means that there is no change in the action. The sizes of the "minbu" and the "parent" remain the same, only the position of the "bottleneck" is changed. The move also preserves the total number of "bottlenecks". Therefore the detailed balance condition is satisfied if the "bottleneck" from which we start the move is chosen randomly from the set of all "bottlenecks" of the manifold.

We found it convenient to organize the "baby universe surgery" moves into a "global sweep". First we find all "bottlenecks" and form their list. The "baby universe surgery" is attempted for a randomly chosen "bottleneck" as many times as is the number of the "bottlenecks". The move is performed *if* the "minbu" size is bigger that a limiting value. In practice this limit was chosen to be 20. After the move the list of the "bottlenecks" is updated. The "global sweep" permits to obtain a distribution of the "minbus" for free, because this information is necessary anyway.

The "baby universe surgery" moves are not ergodic. It is therefore necessary to sup-plement them with the local moves. In the numerical study reported below the "global sweep" is followed by 9 standard sweeps. If the "global sweep" is not included, the autocorrelation times differ very much depending on the type of the measured quantity. Typically one has relatively short autocorrelation times for local quantities, like the average curvature and very long autocorrelation times for the long-range quantities, like the average size of the system. Typically for a system with 16000 simplexes and $k_2 \approx 1.2$ (just below the pseudo-critical value) we get respectively $35 \pm 5$ and $2400 \pm 200$ sweeps. The autocorrelation times for the average size are in this case difficult to be measured, because a typical "time" dependence is so slow and in effect it can easily be underestimated in a short run. Autocorrelation times grow for larger systems as is usually the case (cf. the detailed discussion in the two-dimensional case presented in [37]) and for higher values of the coupling constant $k_2$ when the system enters the branched polymer phase.

The "global sweeps" have small effect on local quantities, but dramatically reduce the autocorrelation time for the non-local quantities. For the system discussed above we get respectively $25 \pm 5$ and $32 \pm 7$ (the "global sweep" is treated as one of the sweeps here). The reduction of the autocorrelation time is even more spectacular in the branched polymer phase, where the measurements of the long-range quantities were practically possible only for small systems. The "baby universe surgery" algorithm permits us to study much larger systems (in our case up to 64000 four-simplexes) near the critical point.

The important question when using the new algorithm is a comparison of the time used by the "global sweep" with that of the standard sweep. For a fixed value of the coupling constant $k_2$ the standard move, defined as $N_4$ local moves, takes time proportional to $N_4$. The performance of the "global sweep" depends on two parameters: the number of "minbus" and the average "minbu" size. Since the distribution of "minbus" is strongly peaked around small sizes the number of performed "baby universe surgeries" will depend on the cut-off in the accepted "minbu" size, which is a parameter of the program. In our calculations we found that the number of performed surgeries can easily be kept proportional to the system size $N_4$. The average "minbu" size reflects the phase structure of the model and is a non-trivial function of $N_4$ and $k_2$ (see fig.2). It shows no $N_4$ dependence in the crumpled phase (small $k_2$) and seems to be proportional to $N_4^{1/2}$ in the so called branched polymer phase (in fact it provides additional evidence of the branched polymer interpretation as we will explain later). In effect the time used for a "global sweep" grows in this phase relatively faster than for a standard sweep. For $k_2 = 1.6$ this relative time increase is only 8/3 between 9K and 64K and even for the largest system the time of the "global sweep" remains of the same order as that of the standard sweep.

## 4.2  Numerical setup

As already mentioned above the details of Monte simulations of four-dimensional simplicial quantum gravity have already been reported elsewhere. The new ingredient in the present simulations, is the baby universe surgery algorithm combined with the increased theoretical understanding of scaling. The algorithm allows us to investigate in detail the elongated phase. We performed simulations for system sizes around 9K, 16K, 32K and 64K systems. By e.g. a 32k system we have in mind triangulations which are constrained to fluctuate in the neighborhood of 32.000 four-simplexes. We cannot keep the number of four-simplexes fixed in the simulations since the local moves in general change this number and they have to be included if the Monte Carlo algorithm is to be ergodic. We follow the method we used in our earlier work and enforce the fluctuations around the chosen size $N_4^0$ by adding to the action the extra term

$$\Delta S = \lambda |N_4 - N_4^0| \qquad (48)$$

with a small parameter $\lambda$ and choosing $k_4$ close to the critical value $k_4^c(k_2)$. The value of $\lambda$ determines the width of the band around $N_4^0$ in which the number of four-simplexes is allowed to fluctuate. A priori it is not known if it is allowed to confine the variation of $N_4$ to a band close to the the desired average value for manifolds which are algorithmically recognizable, but the results reported in [28] support the idea that one can take quite a narrow band of a few thousand, or even less around the average value for all practical purposes. It is anyway possible to check that the results are numerically stable with respect to a given band-width. In practical calculations this width was taken to be 1000.

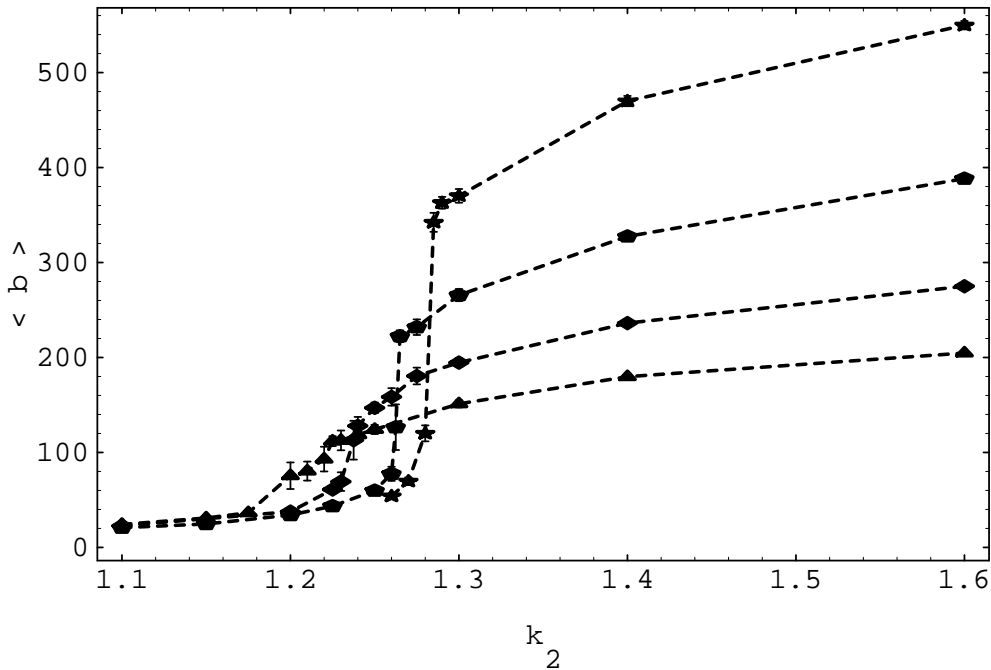

Figure 2: The plot of the average minbu size versus $k_2$ for $N_4 = 9000$ (triangles), 16000 (squares), 32000 (pentagons) and 64000 (stars) simplexes.

In all the measurements we were monitoring the autocorrelation time and the typical run length was 200 to 300 $\tau$ ($\tau$ being the longest autocorrelation time). Near the pseudo-critical point it was sometimes necessary to increase this time to $500\tau$. As was explained in the previous subsection the "baby universe surgery" algorithm permits us to measure the distribution of the "minbu" sizes at the same time when the "global sweep" is performed. Since the number of $N_4$ fluctuates this distribution corresponds in fact to the distribution smeared over the band around $N_4^0$. Since the band-width is small compared to $N_4^0$ we found that this smearing has no effect on the shape of the distribution and permitted us to get very good statistics for the baby universe distribution for free.

The measurements of the distributions of $\langle n(r) \rangle_{N_4}$ are the most time consuming part of the program[8]. An accurate measurement requires averaging over the starting position, in principle over all four-simplexes of the manifold. In practice we found that we can get reasonable accuracy by restricting this averaging to a randomly chosen set containing 10% of the manifold four-simplexes. Below this fraction errors increase rapidly. Even with this reduction the measurement time for a manifold with size $N_4$ grows like $N_4^2$. For the 64K system 80% of the computer time is used by the measurement of $\langle n(r) \rangle_{N_4}$.

## 4.3 The critical point

The first question we want to address concerns the existence of a critical point $k_2^c$ which separates two phases of simplicial quantum gravity in four dimensions. This transition has been known since the first simulations were performed. Recently it has been conjectured that it is a cross over and not a genuine transition [23]. Here we show that this is not the case.

Define the average linear size of a universe of volume $N_4$ in the following way:

$$\langle r \rangle_{N_4} = \left\langle \frac{1}{N_4^2} \sum_{s_4}^{N_4} \sum_r r(s_4) n(r(s_4)) \right\rangle_{N_4}, \qquad (49)$$

where the average is performed over the different triangulations of $S^4$ consisting of $N_4$ four-simplexes which are generated during the simulations. $s_4$ denotes a four-simplex and $n(r(s_4))$ denotes the number of four simplexes in the geodesic distance $r(s_4)$ of $s_4$. The first summation is over all four-simplexes, the next over all distances. For both summations we have a normalization factor $1/N_4$. From the discussion in section 3 we expect in the case of a finite Hausdorff dimension that[9]

$$\langle n(r) \rangle_{N_4} \sim r^{d_H - 1} \ e^{-c\left(r N_4^{1/d_H}\right)^{d_H/(d_H - 1)}}, \qquad (50)$$

i.e. we get:

$$\langle r \rangle_{N_4} \sim \sum_r \frac{r^{d_H}}{N_4} \ e^{-c\left(r N_4^{1/d_H}\right)^{d_H/(d_H - 1)}} \ \sim \ K \ N_4^{1/d_H}, \qquad (51)$$

where the constant $K$ is

$$K = \int dt \ t^{d_H} \ e^{-c \ t^{d_H/(d_H - 1)}}.$$

Eq. (51) is the "fixed $N_4$" version of $\langle N_4 \rangle_r \sim r^{d_H}$.

In fig. 3 is shown a plot of the average radius of universes as a function of the coupling constant $k_2$ for various sizes of the universe ($N_4$= 9K, 16K, 32K and 64K). It is clear that we see two phases, one where $\langle r \rangle_{N_4}$ grows quite rapidly with $N_4$ and one where there is almost no dependence on $N_4$. It is also clear that we observe a drift towards higher values of $k_2$ for the so called pseudo critical coupling $k_2^c(N_4)$. One can define the pseudo critical coupling in various ways. For spin systems one would use the peak in the specific heat or the peak in the spin susceptibility to identify the pseudo critical coupling. In such systems one expects a behavior:

$$|\beta_c(V) - \beta_c| \sim \begin{cases} V^{-1} & \text{for first order transitions} \\ V^{-\frac{1}{\bar{\nu} d}} & \text{for higher order transitions} \end{cases}, \tag{52}$$

where $\bar{\nu}$ is the critical exponent for the spin correlation function (not to be confused with $\nu = 1/d_h$ introduced above) and $d$ the dimension of space.

---

[8]They are performed only for $N_4 = N_4^0$, in contrast to the measurements of the minbu distributions

[9]Strictly speaking this formula is not valid: It is a hybrid of the short distance and long distance behavior which is convenient for deriving scaling relations like (51), but the power correction to the exponential decay is in general different from $r^{d_H - 1}$.

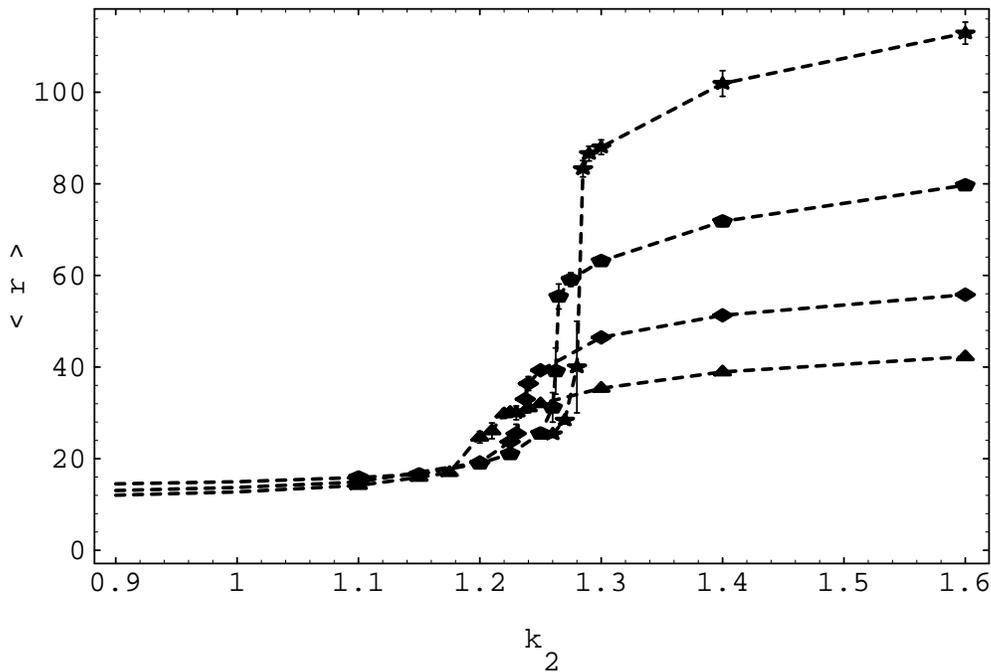

Figure 3: The plot of $\langle r \rangle_{N_4}$ versus $k_2$ for $N_4 = 9000$ (triangles), 16000 (squares), 32000 (pentagons) and 64000 (stars) simplexes.

In the case of gravity it is natural to define the susceptibility as the second derivative of the partition function with respect to $k_2$. In the same way as the magnetic susceptibility of a spin system can be given the interpretation as the integral of the spin-spin correlation function, we have seen that the second derivative of the free energy with respect to $k_2$ has an interpretation as the connected curvature-curvature correlation function:

$$\chi_R(k_2, N_4) = \frac{d^2 \log Z(k_2; N_4)}{d^2 k_2} \sim \left\langle \int d^4\xi_1 d^4\xi_2 \sqrt{g} R(\xi_1) \sqrt{g} R(\xi_2) \right\rangle_c, \tag{53}$$

where we have used a continuum notation. This susceptibility has indeed been used in the former studies. However, the peak is rather asymmetric for reasons to be explained below, and a precise location of the peak is not so clear. We find that the location of the pseudo-critical point $k_2^c(N_4)$ is more precisely determined from a measurement of the critical exponent $\gamma$ discussed in section 3. We will discuss these measurements in detail below. Here it is sufficient to show the best values of $k_2^c(N_4)$ from these measurements in fig. 4.

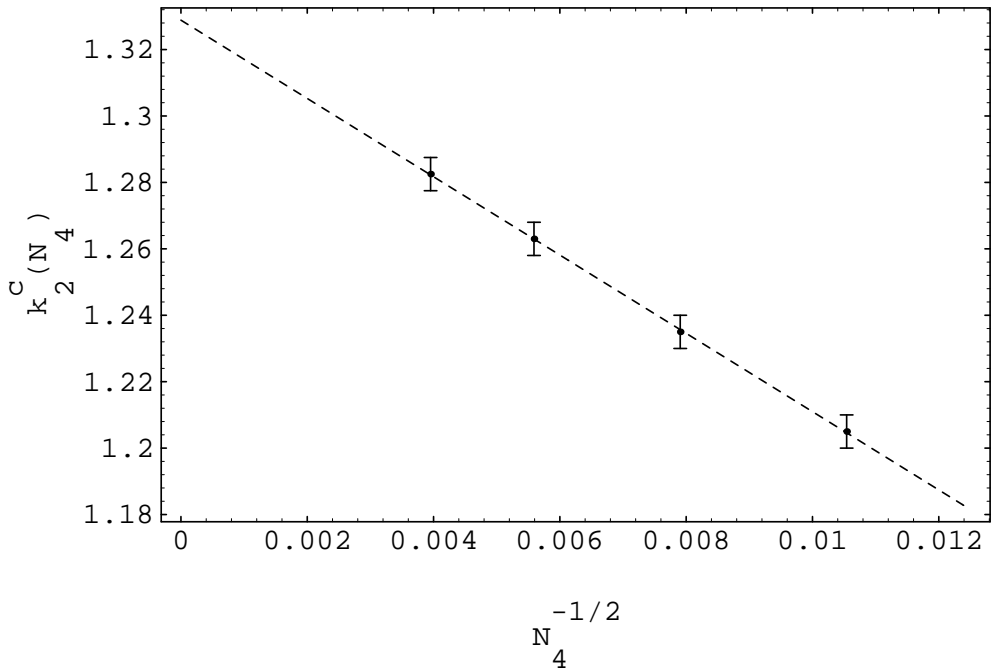

Figure 4: The pseudo critical point $k_2(N_4)$.

We draw two conclusions from fig.4: There *is* an infinite volume extrapolation $k_2^c$ of $k_2^c(N_4)$ and it follows the finite size scaling formula (52):

$$|k_2^c(N_4) - k_2^c| \sim \frac{1}{N_4^\delta}. \tag{54}$$

The best fit is obtained for $\delta = 0.47 \pm 0.03$ with the corresponding critical value $k_2^c = 1.336 \pm 0.006$. It is tempting to conjecture that $\delta = 1/2$ as shown on fig.4. The value of $\delta$ indicates that the transition *is not* a first order transition. In case we use $d = 4$ since we consider four-dimensional manifolds[10] we get from (52) that *the exponent* $\bar{\nu} = 1/2$.

### 4.4 The entropy exponent $\gamma$

We have already defined the entropy exponent $\gamma$ above as the subleading powerlike contribution to the partition function $Z(k_2; N_4)$ or equivalently as the exponent which appears in the most singular part of the second derivative of the partition function with respect to the cosmological constant. A convenient way to measure $\gamma$ is by "minbu counting". Since we are already using a computer algorithm which identifies the baby universes this counting is for free, as already mentioned above. In the elongated phase $k_2 > k_2^c$ there is a pronounced fractal structure and the distribution can be measured with great precision.

---

[10]It should be emphasized that precisely this aspect of finite size scaling is not yet clarified in the context of quantum gravity. For ordinary statistical systems the natural scale which enters in eq.(52) for higher order transitions is the linear dimension for the system. This linear dimension $L$ is usually given to us by the geometry. Here we are integrating over geometries and only the total volume is unambiguously defined. This is why we reformulated (52) to be a function of the volume. The price is that only the combination $\nu d$ appears. What value of $d$ should we choose in the case of quantum gravity? The Hausdorff dimension $d_H$ or the dimension $d$ of the manifold. To our knowledge the answer is not known.

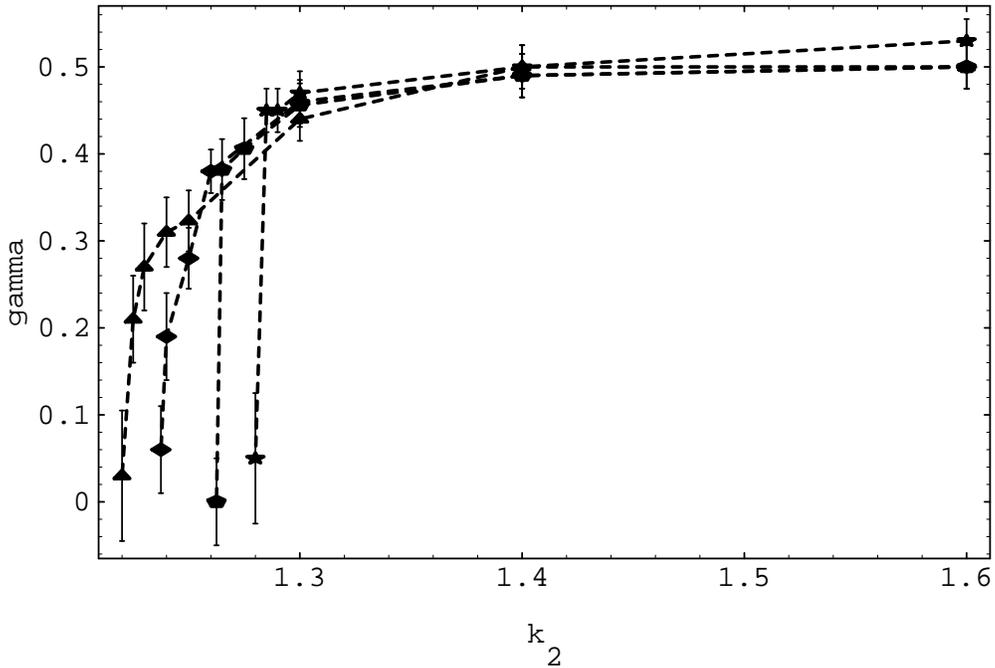

Figure 5: The measure values of $\gamma$ for $k_2 > k_2^c$ for $N_4 = 9000$ (triangles), 16000 (squares), 32000 (pentagons) and 64000 (stars).

We find that $\gamma \approx 1/2$. At this point it should be mentioned that $\gamma = 1/2$ is a theoretical upper limit. The proof of this is identical to the corresponding one for two-dimensional quantum gravity coupled to matter [4]. In fig. 5 we have shown the values of $\gamma$ extracted from the measured distribution of baby universes. We simply fit the measured distribution to

$$\log \mathcal{N}(V_4) = (\gamma - 2)\log[V_4(1 - V_4/N_4)] + \text{const.} \tag{55}$$

and make a lower cut such that only $V_4 > V_4(0)$ is included. Such a lower cut is expected since the theoretical formula is only valid for large $V_4$. Typical cut-off value for $V_4$ is 200. In fig. 5 we show the extracted value of $\gamma$ versus $k_2$. It is seen that $\gamma \approx 1/2$ for $k_2 > k_2^c$ where $k_2^c$ is approximately the value where the geometry changes according to fig. 3. In fact the measurement of $\gamma$ seems to give us the most precise signal for this change, a fact already utilized in fig. 4. Close to the transition $\gamma$ starts to drop and while the geometry in this region fluctuates quite a lot and makes precise measurements difficult, it seems that $\gamma \approx 0$ *at* the transition point. Below the transition it is impossible to extract a $\gamma$ from the baby universe distribution[11] and as we will report below the asymptotic form (36) of the partition function is not valid in this region. It is possible to make a fit to (39) with the asymptotic form (37), and where $\alpha$ seems to change continuously from approximately 1 to 1/4 when $k_2$ decreases from $k_2^c$ to $k_2 = 0$. The precise determination of $\alpha$ was not possible, because acceptable fits are obtained for almost any value in this range.

---

[11] In an earlier letter [19] we reported about such measurements, but also remarked that the data for $k_2 < k_2^c$ did not fit well to the straight line in the log-log plot corresponding to eq. (55) for large $V_4$. We attributed it to the bad statistics of large baby universes, which are sparse in this phase. However, the evidence now points to the non-existence of a $\gamma$ in this phase and in fact a distribution like (36) is disfavored for $k_2 < k_2^c$. We therefore retract the claims in [19] for $k_2 < k_2^c$ concerning $\gamma$.

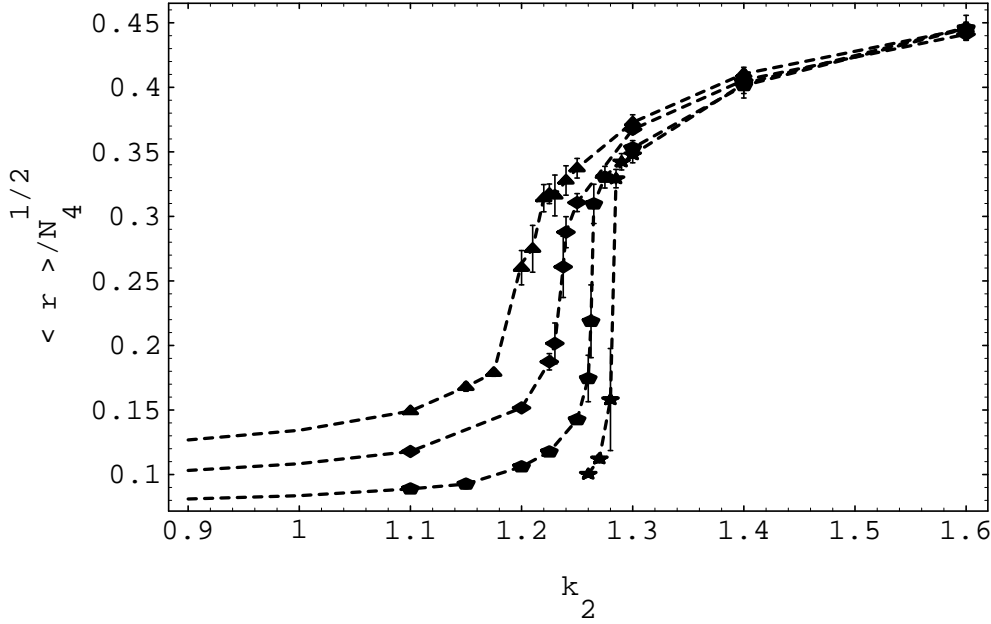

Figure 6: $\langle r \rangle_{N_4} / \sqrt{N_4}$ versus $k_2$ for $N_4 = 9000$ (triangles, 16000 (squares), 32000 (pentagons) and 64000 (stars). For $k_2 = 1.4$ and 1.6, i.e. away from the critical point $k_2 = 1.336$, the values are independent of $N_4$.

In the next subsection we will provide additional evidence that $\gamma = 1/2$ for $k_2 > k_2^c$.

## 4.5 The distribution $\langle n(r) \rangle$

The measurement of $\langle n(r) \rangle_{N_4}$ is easy in the numerical simulations. One simply counts the number of four-simplexes in shells of distance $r$, and one can recursively work one's way out from $r$ to $r + 1$ until the whole manifold is covered. Taking the average over different starting positions and the average over independent triangulations with a fixed $N_4$ will provide us with $\langle n(r) \rangle_{N_4}$.

As discussed in section 3 we expect from theoretical considerations that *if* the ensemble of piecewise linear manifolds has a characteristic internal Hausdorff dimension the function $\langle n(r) \rangle_{N_4}$ should obey the scaling law:

$$n(r, N_4) \equiv \frac{1}{N_4} \langle n(r) \rangle_{N_4} = N_4^{\frac{1}{d_H}} \, n(r/N_4^{1/d_H}). \qquad (56)$$

In addition we expect that the short distance behavior of $n(x)$ to be like $x^{d_H - 1}$ and the long distance behavior to be like $\exp(-c \, x^{d_H/(d_H-1)})$.

In fig. 6 we have tested the scaling behavior of $n(x)$ for $\langle r \rangle_{N_4}$ and $k_2 > k_2^c$. If the form (56) is satisfied $\langle r \rangle_{N_4} = \sum_r r \, n(r, N_4) \sim N_4^{1/d_H}$. In the figure $d_H = 2$ and it is clear that we have perfect scaling for $k_2$ not too close to $k_2^c(N_4)$, i.e. for $k_2$ above $k_2^c = 1.34$ there should be scaling for all values of $N_4$.

The data are sufficiently good that we can in fact fit to a detailed functional form for $n(x)$. For $k_2 = 1.29$ and $N_4 = 64.000$ we show in fig. 7 the data and the function:

$$n(x) \sim x \, e^{-cx^2}. \qquad (57)$$

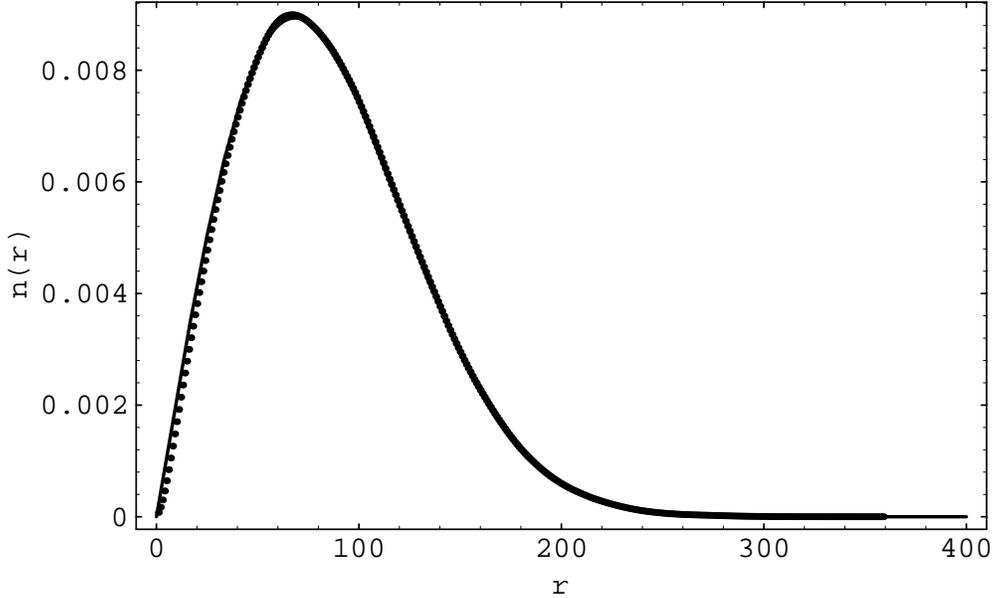

Figure 7: The distribution $n(r, N_4)$ for $N_4 = 64.000$ and $k_2 = 1.29$. The dots are the measured values (error bars less than the dots) while the curve is the best fit of the form (57).

The data points are the dots (the error bars are less than the dots) and the curve is just $n(x)$ from (57) with only one adjustable parameter ($c$). It is seen that the agreement is almost perfect in spite of the fact that $k_2 = 1.29$ is just above the transition point (the pseudo critical point for $N_4 = 64000$). If we move further into the elongated phase the agreement gets even better. This implies in particular that the distributions for the same $k_2$ can be scaled on top of each other by using $x = r/\sqrt{N_4}$ in this phase. In fact the agreement is so good that it would be impossible to distinguish systems of different sizes for $k_2 = 1.4$ and $1.6$.

The conclusion so far is that the elongated phase $k_2 > k_2^c$ is characterized by $\gamma = 1/2$ and $d_H = 2$ and that

$$G(r; N_4) \sim e^{k_4^c(k_2)N_4} N_4^{-3/2} r e^{-cr^2/N_4} \tag{58}$$

From this we can construct the two-point function:

$$G(r; k_4) = \sum_{N_4=1}^{\infty} G(r; N_4) e^{-k_4 N_4} \sim e^{-\tilde{c}\sqrt{\Delta k_4} r}. \tag{59}$$

*This two-point function is precisely the two-point function of a theory of branched polymers [4] and the picture which emerges is that the typical manifolds in this region of the coupling constant space are like small tubes which can join and branch in tree-like structures.*

The transition away from the branched polymer region takes place increasingly abruptly with increasing $N_4$ and we conjecture that the phase in the infinite volume limit continues all the way down to $k_2^c$. In fact this picture would be consistent with the observed finite size scaling exponent $\bar{\nu} = 1/2$ introduced below eq. (54), if we identify this exponent with the mass exponent of the two-point function which is $\nu = 1/d_H$. *At* the transition point $k_2^c$ there could be a different critical behavior. It is quite difficult to perform measurements exactly at the transition. The geometry fluctuates a lot, but as already mentioned the minbu counting suggests $\gamma \approx 0$. Measurements of $d_H$ suggests $d_H \approx 4$, but we must insist that the data do not allow a clear extraction of $d_H$.

The measurements of $\langle n(r) \rangle_{N_4}$ for $k_2 < k_2^c$ show completely different characteristics than for $k_2 > k_2^c$. For large $r$ there is *no* $N_4$ dependence at all and the function seems to fall off exponentially:

$$\langle n(r) \rangle_{N_4} \sim \mathrm{e}^{-m(k_2)\,r}. \tag{60}$$

As discussed above such a behavior is consistent with the notation of infinite Hausdorff dimension. However, if this notation should be "genuine" it should also reflect itself at small distances, i.e. contrary to the situation for smooth four-dimensional manifolds with negative curvature it should not be possible to have $\langle n(r) \rangle_{N_4} = r^3$ for *small* $r$. We have not been able to fit to any sensible power dependence $r^{d_H-1}$ for small $r$. A plot of $\log \langle n(r) \rangle_{N_4}$ shows indeed a linearly growing function of $r$ up to some $r_0(N_4)$ which can be identified with both the peak of the distribution $\langle n(r) \rangle_{N_4}$ and with the average value $\langle r \rangle_{N_4}$. $r_0(N_4)$ or $\langle r \rangle_{N_4}$ shows only a very weak dependence of $N_4$ and a reasonable fit to $\langle r \rangle_{N_4}$ is

$$\langle r \rangle_{N_4} = a(k_2) + b(k_2) \log N_4. \tag{61}$$

This again gives some support to the idea that the Hausdorff dimension is infinite in this phase. Recall from section 3 that an infinite Hausdorff dimension will result in an exponential dependence like in (60) if it appears as a limit of distributions with finite Hausdorff dimension. Similarly eq. (61) would appear as the limit of $\langle r \rangle_{N_4} \sim N^{1/d_H}$ for $d_H \to \infty$.

The observation that $\langle n(r) \rangle_{N_4}$ grows exponentially from $r \approx 6$ out to $r \approx r_0$ and then falls off exponentially indicates that we deal with an infinite Hausdorff dimension at all distances and it is easy to get a quite good "phenomenological" fit to $\langle n(r) \rangle_{N_4}$ which incorporates both these features by choosing e.g.:

$$\langle n(r) \rangle_{N_4} \sim \exp\left( -m_1(k_2)r - c_2 \mathrm{e}^{-m_2(k_2)r} \right). \tag{62}$$

It will grow like $\mathrm{e}^{(c_2 m_2 - m_1)r}$ for small distances and fall off like

$$\langle n(r) \rangle_{N_4} \sim c_1\, \mathrm{e}^{-m_1(k_2)\,r} - c_2\, \mathrm{e}^{-(m_1(k_2)+m_2(k_2))\,r} + \dots \tag{63}$$

for large distances, while a $N_4$ dependence in the coefficient $c_2$ would explain the observed $N_4$ dependence of $r_0$.

The data and a fit of the form (62) are shown in fig. 8 for $k_2 = 1.26$ and $N_4 = 64000$, i.e. right below the transition to the crumpled phase, where the fit is worst. But even so close to critical point (62) works quite well over the whole range of $r > 6$. It should be mentioned that the coefficient in front of the second exponential in eq. (63) is negative. This implies that the term cannot be given the interpretation as an additional heavier mass excitation. However, just looking at the long distance tail the distribution $\langle r \rangle_{N_4}$ allow us to determine $m_1$, $m_2$ and $c_2$ from (63). On the other hand we can determine $c_2 m_2 - m_1$ from the short distance exponential growth alone and find good agreement. This indicates that long and short distance behavior are intertwined in the case of infinite Hausdorff dimension, as they are in the case of finite Hausdorff dimension, where $d_H$ appears both in the short distance and long distance expression for $\langle r \rangle_{N_4}$ (see e.g. (50)).

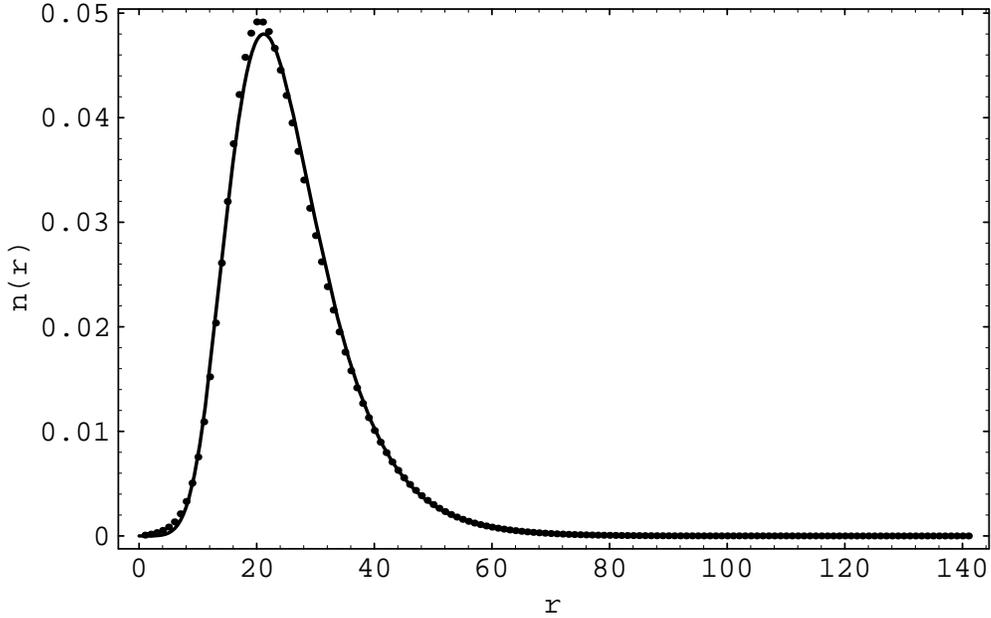

Figure 8: exponential fit (curve) of the form (62) to the measured $n(r; N_4)$ (dots, error bars less than dot-size) in the crumpled phase ($N_4 = 64000$, $k_2 = 1.26$).

*We conjecture that the internal Hausdorff dimension is infinite for $k_2 < k_2^c$.*

It is worth to emphasize that (60) is inconsistent with even a logarithmically weak scaling of the mass of the two-point function $G(r; k_4)$ for $\triangle k_4 \to 0$. As mentioned in section 3 it would require at least a logarithmic dependence like $e^{-r/\log(N/r)}$. Our data seem not to favor such dependence and this implies that the two-point function (by the usual discrete Laplace transformation) will be:

$$G(r; k_2, k-4) \sim A_1 e^{-m_1(k_2) r} + A_2 e^{-(m_1(k_2)+m_2(k_2) r)} + \cdots, \tag{64}$$

The $m_i(k_2)$'s do not scale to zero as $k_4 \to \kappa_4^c(k_2)$, but in fig. 9 we have shown their limiting value at $k_4^c(k_2)$ as a function of $k_2 - k_2^c(N_4)$. *The data seem to fall on universal curves and the lowest mass $m_1(k_2)$ scales to zero as $k_2 \to k_2^c$.* If we extract a critical exponent $\tilde{\nu}$ (not to be confused with the former introduced mass exponents $\nu = 1/d_h$ and $\bar{\nu}$ from finite size scaling) for $m_1(k_2)$ we get:

$$m_1(k_2) \sim |k_2 - k_2^c|^{\tilde{\nu}}, \quad \tilde{\nu} = 0.50 \pm .01. \tag{65}$$

*The fact that we have a mass excitation which scales to zero at the critical point $k_2^c$ is a strong indication that the system will develop genuine extension and a finite Hausdorff dimension at the critical point.*

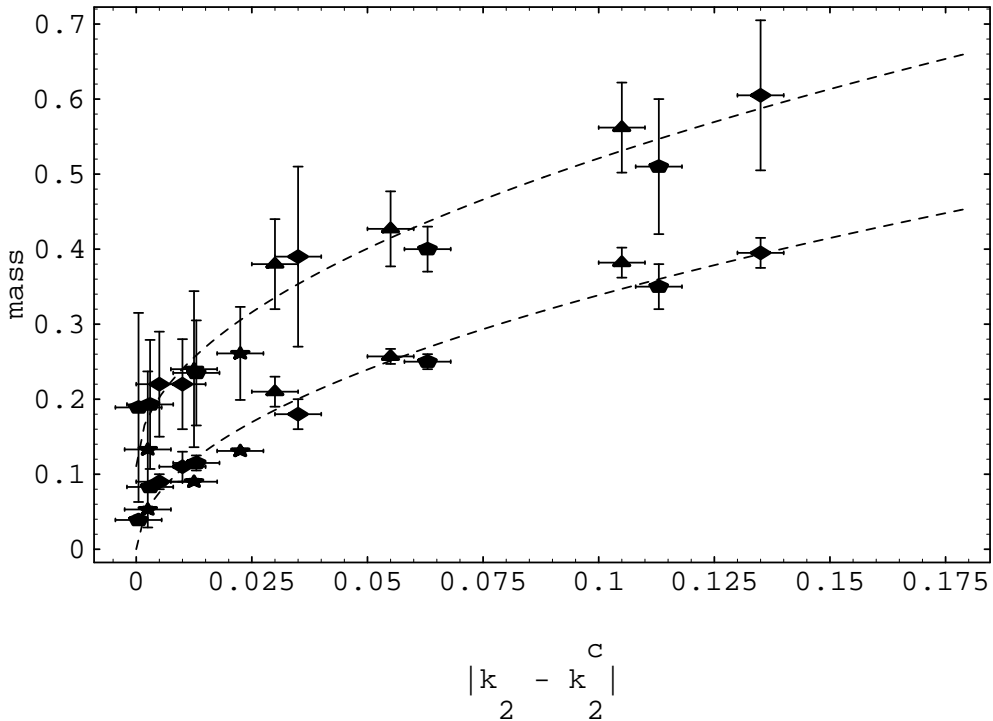

Figure 9: The behavior of the masses $m_1(k_2)$ and $m_1(k_2) + m_2(k_2)$ from (62) in the crumpled phase as a function of $k_2^c - k_2$.

## 4.6 Miscellaneous

Let us collect here some of the observations which were scattered in the presentation of the data, but which did not fit into the discussion earlier.

The first observation is that in the elongated phase the average "minbu" size grows as $N_4^{1/2}$, $N_4$ being the total volume of the universe. This was mentioned above and is shown in fig. 10. Is it understandable? The answer is yes if we assume that $\gamma = 1/2$. If we assume there exists a $\gamma$ at all we calculated earlier the probability distribution for minbu universes of size $V_4$ to be $N_4 \left[V_4(1 - V_4/N_4)\right]^{\gamma-2}$ (see eq. (40)). The average size is therefore

$$\langle V_4 \rangle_{N_4} = \sum_{V_4=1}^{N_4/2} V_4 \, N_4 \left[V_4(1 - V_4/N_4)\right]^{\gamma-2} \approx N_4^\gamma \int_0^{1/2} dt \, t^{\gamma-1}(1-t)^{\gamma-2} \qquad (66)$$

if we assume $\gamma > 0$. The data support the value $\gamma = 1/2$.

The next observation concerns the behavior of the pseudo-critical point $k_2^c(N_4)$ as a function of $N_4$. According to eq. (54) it approaches the true critical point $k_2^c$ as $1/\sqrt{N_4}$ and we conjectured that it corresponded to critical exponent $\nu = 1/2$. Do we have such mass exponents around as a function of $k_2$ ? The answer is yes: The mass $m_1(k_2)$ in the crumpled phase seems go to zero as $|k_2 - k_2^c|^{\tilde{\nu}}$, where the exponent $\tilde{\nu} \approx 1/2$. A closer analysis of the data of for instance the average curvature and the susceptibility $\chi_R(k_2, N_4)$ indicates that the major part of the finite size effect comes from the crumpled phase. The curve corresponding to $\chi_R(k_2, N_4)$ shows a clear asymmetry as a function of $k_2$ around the peak at $k_2^c(N_4)$. It grows relatively slowly from the crumpled phase when $k_2 < k_2^c(N_4)$ approaches $k_2^c(N_4)$ and drops abruptly for $k_2 > k_2^c(N_4)$. The asymmetry can now be partly understood: In the crumpled phase we have no masses scaling to zero as $k_4 \to k_4^c(k_2)$. Only when $k_2$ approaches $k_2^c$ will the lowest mass scale to zero. In the elongated phase we *always* have a "mass" excitation $m(k_4)$ which scales to zero as $k_4 \to k_4^c(k_2)$ like $(k_4 - k_4^c(k_2))^{1/2}$. This true for all values of $k_2 > k_2^c$. It is natural to conjecture that the finite size dependence of the pseudo-critical point $k_2(N_4)$ is monitored by the mass $m_1(k_2)$ in the crumpled phase.

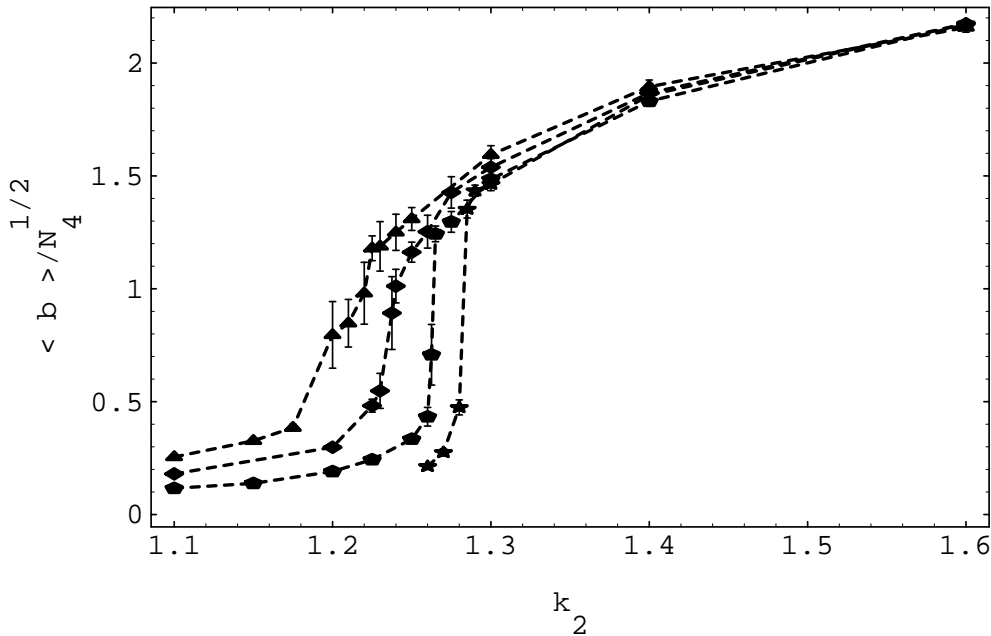

Figure 10: The scaled average minbu size $\langle b \rangle_{N_4}$ defined in eq.(66) versus $k_2$.

## 5  Discussion

### 5.1  Summary of observations

Four dimensional simplicial quantum gravity as defined in this article depends on two bare coupling constants, $k_2$ and $k_4$. The first serves as the inverse of the bare gravitational coupling constant while the second one in the formal continuum limit is a linear combination of the inverse gravitational coupling constant and the cosmological constant. We expect at least $k_4$ to have an additive renormalization, i.e. the continuum limit of the theory should be defined as a function of $k_4 - k_4^c(k_2)$. In the $(k_2, k_4)$ coupling constant plane there is a critical line $k_4 = k_4^c(k_2)$ where such a limit can be taken. This leads to an infinite (lattice) volume of the theory. The renormalized cosmological constant will be defined as a function of $k_4 - k_4^c(k_2)$. It might not be possible to define an interesting continuum field theory for all values of $k_2$. Recall the analogous situation for a spin model, e.g. the Ising model, where there is only a single point where a continuum limit can be taken.

We have observed two distinct phases of the discretized theory. One phase, denoted the elongated phase, was characterized by specific scaling properties of the two-point function, which allowed us to conjecture that the Hausdorff dimension of the ensemble of (piecewise linear) four-manifolds considered was $d_H = 2$ while the entropy exponent $\gamma$, which determines the proliferation into baby universes, was $\gamma = 1/2$. The detailed investigation of the geometry in this phase showed perfect agreement with a model of fractal structures denoted *branched polymers*. This phase was present as long as the inverse bare gravitational coupling constant $k_2 > k_2^c$, i.e. for small ("semiclassical") values of the bare gravitational coupling constant.

For $k_2 < k_2^c$ we have a completely different phase. From the numerical data we were led to conjecture that the Hausdorff dimension $d_H$ is infinite and that there is no entropy exponent $\gamma$, in agreement with recent results obtained for $k_2 = 0$ [28,29]. In this phase we can say that there is effectively no metric associated with the volume: The volume enclosed in a geodesic ball of radius $r$ grows exponentially with $r$, but contrary to the situation for smooth spaces with negative curvature this is true down to the smallest distances of $r \approx 3 - 4$. This implies that the main part of a typical universe can be reached in few steps, and we can view the universe to be of Planck size. This is possible because there exist vertices of very high order in the typical triangulation, and such vertices ensure that a large part of the universe can be connected in few steps. Nevertheless the exponential decay of the two-point function allows us to identify a mass which scales to zero when $k_2 \to k_2^c$. This strongly suggests that the ensemble of four-manifolds will develop a genuine extension when $k_2 \to k_2^c$.

It would be very interesting if we could determine the precise value of the Hausdorff dimension *at* the critical point $k_2^c$, as well as the entropy exponent $\gamma$ if it exists in this point. The present data hint that $\gamma(k_2^c) \approx 0$ but it is clear from the curves of fig. reffig4 that the correct should rather be: in the critical region, where $\gamma$ changes very fast and probably even jump from $\gamma = 1/2$, we have measured values of $\gamma \approx 0$, but never values below zero. This does not rule out that increased fine tuning of $k_2 \to k_2^c$ from above would allow us to reach negative values of $\gamma$.

## 5.2   Comparison with Regge discretization

Apart from the dynamical triangulation approach to quantum gravity, there exists another approach which here we shall denote Regge quantum gravity (RQG). The approach is completely different in philosophy from simplicial gravity. While simplicial gravity should be viewed as a discretization directly in the space of equivalence classes of metrics, RQG works with a fixed triangulation and varies the length of the links. In this way many configurations might correspond to the same metric and in principle a Jacobian should be included in the functional integral. This may be a serious problem for the approach. On the other hand it has the advantage that it allows for a weak coupling expansion which can be important if we want to relate to a continuum formulation. For an review we refer to [41]. Recently some problems have been reported both in two-dimensional gravity where the simplest choices of measures do not reproduce the known continuum results [38,39] and in four dimensions, where a drastic dependence on the choice of the fixed triangulation used in the computer simulations was observed [40]. If we ignore these problems and just refer to the results as reported in [41] there seems to be some similarity in the results obtained from RQG and the ones reported here. First of all there are two phases in both approaches. The elongated phase is not particularly well defined in RQG, but it still has some flavor of the elongated phase seen in simplicial quantum gravity.

In the other phase a mass gap is observed, precisely as in our simulations and the mass scales to zero with approximately the same exponent as $\bar{\nu}$ in eq.(65). There seem to be divergent opinions concerning the *sign* of the volume-volume correlation function. In [42] it is reported negative, while in [43] positive. In simplicial gravity the correlation function is by definition positive.

There are also differences in the two approaches. In simplicial quantum gravity the naively defined average scalar curvature does not scale to zero at the critical point. It is not presently known if this is due to the fact that the curvature operator is a composite operator which requires a subtraction before contact can be made with continuum. In that case the average value of the curvature should be absorbed in a redefinition of the cosmological constant. Whatever is the explanation, it disagrees with the measurements from RQG where the average curvature is seen to scale to zero when $k_2 \to k_2^c$. Another, more drastic difference is indicated by the Hausdorff dimension in the strong coupling phase (i.e. the crumpled phase in simplicial quantum gravity). As reported above it seems to be infinite in crumpled phase. In RQG it is reported to be close to four.

## 5.3   Continuum interpretation

As we have shown above simplicial quantum gravity is a perfectly well defined statistical theory where it is possible to define and discuss the continuum limit and derive scaling relations in a way similar to what is done in the ordinary theory of critical phenomena. Here, as in statistical mechanics, it might be a problem to identify a continuum theory which corresponds to the observed critical phenomena. The problem is that we are presumably far away from any "naive" continuum limit. This is not necessary an undesirable situation. Any "naive" continuum limit of Euclidean quantum gravity is a disaster because the action is unbounded from below. A number of cures have been suggested: The addition of suitable higher derivative terms can bound the action and at the same time make the theory renormalizable [44]. A sophisticated choice of functional measure starting from the formal expression in spaces with Minkowskian signature can also tame the unboundedness [45] (see [46] for a recent review). Finally the problems with the unboundedness of the action should disappear in a topological phase of quantum gravity, where the metric should play no role. Three-dimensional quantum gravity in the formulation of Witten [47] provides an instructive example of how a "minor" change from metric to drei-beins allows us to view the theory as topological and how the rotation to Euclidean signature results in a bounded action. Further the Euclidean theory has a beautiful discretized realization, the so called Turaev-Viro theory [48], which in spirit is not far from simplicial quantum gravity. In fact it is presently not ruled out that three-dimensional quantum gravity in the crumpled phase can be given a representation as topological gravity.

In simplicial quantum gravity we have no problems with an unbounded action as long as the lattice volume is finite. However, part of the regularization is performed directly in the equivalence classes of metrics. This makes it difficult to decide which of the above scenarios is realized by the regularization, and it could depend on the choice of the bare (inverse) gravitational coupling constant $k_2$. Due to the lack of analytic solutions of the discretized theory as in two dimensions, we have to take the numerical data and look for scaling behavior which can be associated with one of the above continuum models.

In the elongated phase we understand now in detail the statistical theory, namely the one of branched polymers. It can be viewed as the ultimate dominance of volume-volume fluctuations and therefore of the conformal factor. This makes it natural to compare with the models of four dimensional quantum gravity which study the conformal sector based on the trace anomaly [49,50]. This approach, which we denote conformal four-dimensional quantum gravity (CFQG), is inspired by the corresponding treatment in two dimensions, but the same is true for our generalization of the successful two-dimensional simplicial quantum gravity theory. CFQG studies the fluctuations of the conformal factor $e^{2\sigma(x)}$ in space-times of the form:

$$g_{\mu\nu} = e^{2\sigma(x)}\bar{g}_{\mu,\nu}(x),$$ (67)

where $\bar{g}_{\mu\nu}(x)$ is a fiducial metric. The theory has an infrared fixed point and at this fixed point the field $\sigma$ receives quantum dressing. The effective action can be written as [51]

$$S_{eff} = \int d^4\xi \sqrt{\bar{g}} \left\{ \frac{Q^2}{16\pi^2} \left[ \sigma\Delta_4\sigma + \frac{1}{2}\bar{G}\sigma \right] - \frac{3}{\kappa} \left[ \alpha^2(\partial\sigma)^2 e^{2\alpha\sigma} - \frac{1}{\kappa} f(\frac{\alpha^2}{Q^2}) e^{4\alpha\sigma} \right] \right\}$$ (68)

In this formula the first term is the trace anomaly induced action and $\kappa = 8\pi G$, while the second term is the Einstein action action. $\alpha$ denotes the quantum gravitational dressing coefficient, $\bar{\Delta}_4$ a fourth order differential operator with respect to the fiducial metric $\bar{g}$, $\bar{G}$ the Gauss-Bonnet term, while $Q^2$ is a kind of analogue of the two-dimensional central charge. Finally $f(x) = x(1+4x+6x^2)$. In addition the partition function at fixed volume will scale like [51]:

$$Z(\kappa, V) \sim V^{\gamma-3} Z\left(\kappa V^{1/2}\right), \quad \gamma = 2 - \frac{Q^2}{2\alpha}.$$ (69)

The dressing coefficient $\alpha$ is related to the gravitational "central charge" by:

$$\alpha = \frac{1 - \sqrt{1 - 4/Q^2}}{2/Q^2} = \frac{2(2-\gamma)}{3 - 2\gamma}.$$ (70)

In particular we expect from (69) that the approach to the infinite volume limit is:

$$\kappa \sim \frac{1}{V^{1/2}}, \quad \text{i.e.} \quad k_2 - k_2^c \sim \frac{1}{N_4^{1/2}},$$ (71)

where the last equation is the translation to the discretized notation of simplicial quantum gravity and where we have identified the critical point $k_2^c$ observed in simplicial quantum gravity with the infrared fixed point of CFQG.

We are now in a position to compare with the "numerical observations" reported above. At the critical point we found $|k_2^c(N_4) - k_2^c| \sim N_4^{-\delta}$, where $\delta = 0.47 \pm 0.03$. The finite size dependence of the pseudo critical point is derived from finite size scaling, where the assumption is that the (singular part with respect to $k_2$) of the free energy $F(k_2, N_4) = \log Z(k_2, N_4)$ can be written as a function $F(N_4^\delta|k_2 - k_2^c|)$. A glance at (71) or (69) shows that the measured $\delta$ indeed agrees with the prediction of CFQG. Unfortunately it is difficult to measure $\gamma(k_2^c)$ very precisely *at* the critical point. Our measurements indicated $\gamma \approx 0$ which corresponds to a value $Q^2 = 16/3$ and $\alpha = 4/3$. According to the hypothesis of "infrared conformal dominance" [51] this value of $Q^2$ should arise by

integrating out the transverse gravity mode, thereby creating an effective action of the same form as (68) when expanded in powers of derivatives.

However, there is some ambiguity in this comparison since it is possible to add an independent volume term to (68) [46,52]:

$$S_{eff} \rightarrow S_{eff} + \Lambda \int d^4 \sqrt{g} \ e^{4\beta\sigma}.$$ (72)

In this case the scaling is changed to

$$Z(\kappa, V) \sim V^{\gamma-3} Z\left(\kappa V^{\alpha/2\beta}\right), \quad \gamma = 2 - \frac{Q^2}{2\beta},$$ (73)

where

$$\beta = \frac{1 - \sqrt{1 - 8/Q^2}}{4/Q^2}, \quad \alpha = \frac{1 - \sqrt{1 - 4/Q^2}}{2/Q^2}.$$ (74)

and we can eliminate the unknown $Q^2$:

$$\beta = \frac{2 - \gamma}{1 - \gamma}, \qquad \frac{\alpha}{2\beta} = \frac{1}{2}(2 - \gamma - \sqrt{1 + (1 - \gamma)^2}).$$ (75)

If we compare (73) with the observed finite size scaling it follows that we only obtain the observed $\alpha/2\beta = 1/2$ if $\gamma \rightarrow -\infty$ at $k_2^c$. As remarked above this can not be ruled out from our measurements. This would correspond to the "semiclassical" region of CFQG where $Q^2 \rightarrow \infty$ and the analogous regime in two-dimensional quantum gravity would be $c \rightarrow -\infty$. If we take $\gamma \approx 0$ we get a value $\alpha/2\beta = 0.29$ which is below the measured value. It is very unfortunate that we presently are unable to determine $\gamma$ more precisely *at the critical point*.

When we move *into* the elongated phase of simplicial quantum gravity it seems that we have reached an extreme limit. In the framework of extended CFQG defined by (72) one would be tempted to view this as the transition where $\gamma \rightarrow 0$ and $\beta$ ceases to be real for $Q^2 \rightarrow 8$. This transition has a flavor of the $c = 1$ barrier in two-dimensional quantum gravity and it has often been conjectured that $c = 1$ (where $\gamma = 0$ in two-dimensional gravity) marks a transition to branched polymers.

To summarize: The comparison of simplicial quantum gravity and $CFQG$ is not entirely unambiguous, partly due to the numerical uncertainty associated with the location of the critical point and partly due to the approximation which is inherent in CFQG by assuming that *all* effects of the transverse degrees of freedom can be absorbed into $Q$. However the qualitative agreement is very encouraging.

In the crumpled phase the situation is more controversial. Seemingly the Hausdorff dimension is infinite (or at least quite large). In addition it has not been possible to define an entropy exponent $\gamma$ and formally $\gamma = -\infty$. This is in agreement with "observations". In the crumpled phase we observe only few and small baby universes. If we move deep into the crumpled phase we find in addition that the naively defined curvature is negative. Does this imply that we observe smooth manifolds with negative curvature? On such manifolds one would expect the volume to grow exponentially with the radius of geodesic balls. The answer seems to be no. Rather the scenario is one where a few vertices have very high order and are linked to almost all other vertices in the manifold. On such piecewise linear manifold the link distance between any two vertices will be only a few lattice units[12] and we see only a very weak dependence of the average radius on the total volume. Nevertheless we know from theoretical considerations that we can define a mass by the exponential decay of the two-point function and from the numerical simulations there are indications of a hierarchy of masses. It is tempting to conjecture that the concept of volume and consequently of metrics cannot play a major role in this phase, even in spite of the ability to define a two-point function. In this case we approach a situation where the linear extension of any universe of finite volume is of Planckian size and where a more adequate description might lead to a theory which has reference only to the topology of the underlying manifold. It is interesting to note that if we choose $k_2 = 0$, i.e. we perform a pure counting of manifolds with no additional weight, we are deeply into the crumpled phase: *the generic piecewise linear manifold is of Planck scale.* As we move towards the semiclassical region (the bare gravitational coupling constant $1/k_2 \to 0$) the ensemble of manifolds undergoes a phase transition: Above the transition the ensemble has a finite Hausdorff dimension $d_H$. In fact $d_H = 2$. It is of course a most interesting question to determine the Hausdorff dimension *at* the transition point. This point seems a candidate for defining a continuum non-perturbative theory of quantum gravity with a finite Hausdorff dimension larger than two and it has the tantalizing interpretation as transition point between the entropically favored Planckian size quantum universe and the two-dimensional fractal universe dominated by the conformal mode.

## 5.4 Outlook

It is worth to recall that the action used in simplicial quantum gravity is extremely simple. We consider this as a virtue of the theory. It has certainly been very useful in the studies of two-dimensional quantum gravity, and there is hope that it will allow us to solve the four-dimensional theory analytically. Until then numerical simulations can be quite useful, serving as inspiration for analytic approaches, and powerful enough to give us important hints about the solution of the theory. The indications that the elongated phase is a branched polymer phase[13] and that the crumpled phase might correspond to a topological phase corroborate on the hope that it might be possible to solve the model analytically.

As already emphasized above it seems very important to develop better algorithms in order to study in detail the neighbourhood of transition point. The baby universe surgery algorithm has been crucial for this numerical study, but there is still room for improvement.

A most interesting question which we have not touched upon is that of topology changes. Hopefully the full power of dynamical triangulations will show us a way to deal with these changes in a quantitative way, as happened in two-dimensional quantum gravity.

---

[12]The corresponding "four-simplex distance" is in general somewhat larger, but it exhibits the same scaling dependence as the link distance.

[13]It should be emphasized that the branched polymer phase is *much* clearer in four-dimensional space than in two-dimensional gravity coupled to matter.

**Acknowledgment** It is a pleasure to thank Ignatios Antoniadis, Emil Mottola and Yoshiyuki Watabiki for interesting discussions and helpful suggestions and Wolfgang Beirl for informing us about the latest results in quantum Regge theory.

# References

[1] F. David, Mod.Phys.Lett. A3 (1988) 1651; J. Distler and H. Kawai, Nucl.Phys. B321 (1989) 509.

[2] F. David, Nucl.Phys. B257 (1985) 45; Nucl.Phys. B257 (1985) 543.

[3] J. Ambjørn, B. Durhuus and J. Fröhlich, Nucl.Phys. B257 (1985) 433; B270 (1986) 457; B275 (1986) 161-184.

[4] J. Ambjørn, B. Durhuus J. Fröhlich and P. Orland, B270 (1986) 457; B275 (1986) 161.

[5] V.A. Kazakov, I. Kostov and A.A. Migdal, Phys.Lett. B157 (1985) 295; Nucl.Phys. B275 (1986) 641.

[6] J. Ambjørn,*Recent progress in the theory of random surfaces and simplicial quantum gravity*, NBI-HE-94-55, Bulletin board hep-lat@ftp.scri.fsu.edu-9412006.

[7] C.F. Baillie and D.A. Johnston, Phys.Lett. B286 (1992) 44.
S. Catterall, J. Kogut and R. Renken, Phys.Lett. B292 (1992) 277.
J. Ambjørn, B. Durhuus, T. Jonsson and G. Thorleifsson, Nucl.Phys B398 (1993) 568.
J. Ambjørn and G. Thorleifsson, Phys.Lett.B323 (1994) 7.

[8] J. Ambjørn, B. Durhuus and T. Jonsson, Mod.Phys.Lett.A9 (1994) 1221.

[9] B. Durhuus, Nucl.Phys. B426 (1994) 203.

[10] M. Wexler, Phys. Lett. B315 (1993) 67; Mod. Phys. Lett. A8 (1993) 2703; Nucl. Phys. B410 (1993) 337; *From trees to galaxies: The Potts model on a random surface*, NBI-HE-94-28.

[11] M.G. Harris and J. Wheater, Nucl.Phys. B427 (1994) 111.

[12] J. Ambjørn and B. Durhuus and T. Jonsson, Mod.Phys.Lett. A6 (1991) 1133.

[13] M.E. Agishtein and A.A. Migdal, Mod. Phys. Lett. A6 (1991) 1863.
B. Boulatov and A. Krzywicki, Mod.Phys.Lett A6 (1991) 3005.
J. Ambjørn and S. Varsted, Nucl.Phys. B373 (1992) 557.
J. Ambjørn, D.V. Boulatov, A. Krzywicki and S. Varsted, phys.Lett. B276 (1992) 432.
N. Sakura, Mod.Phys.Lett. A6 (1991) 2613.
N. Godfrey and M. Gross, Phys.Rev. D43 (1991) R1749.
S. Catterall, J. Kogut and R. Renken Phys.Lett.B342 (1995) 53.

[14] J. Ambjørn and J. Jurkiewicz, Phys.Lett. B278 (1992) 50.

[15] M.E. Agishtein and A.A. Migdal, Mod. Phys. Lett. A7 (1992) 1039.

[16] J. Ambjørn, J. Jurkiewicz and C.F. Kristjansen, Nucl.Phys. B393 (1993) 601.

[17] M.E. Agishtein and A.A. Migdal, Nucl.Phys. B385 (1992) 395.

[18] J. Ambjørn, Z. Burda, J. Jurkiewicz and C.F. Kristjansen, Phys.Rev. D48 (1993) 3695.

[19] J. Ambjørn, S. Jain, J. Jurkiewicz and C.F. Kristjansen, Phys.Lett. B305 (1993) 208.

[20] B. Bruegmann, Phys. Rev. D47 (1993) 3330.

[21] B. Bruegmann and E. Marinari, Phys. Rev. Lett. 70 (1993) 1908.

[22] B. V. De Bakker and J. Smit, Phys.Lett. B334 (1994) 304.

[23] B. V. De Bakker and J. Smit, *Curvature and scaling in 4D dynamical triangulation*, preprint ITFA-94-23. hep-lat/9407014.

[24] S. Catterall, J. Kogut and R. Renken, phys.Lett. B328 (1994) 277.

[25] D. Weingarten, Nucl.Phys. B210 (1982) 229.

[26] W.T. Tutte, Can.J.Math. 14 (1962) 21. F.A. Bender and E.R. Canfield, *The asymptotic number of rooted maps on a surface*, 1986.

[27] S. Catterall, J. Kogut and R. Renken Phys. Rev. Lett. 72 (1994) 4062.

[28] J. Ambjørn and J. Jurkiewicz, Phys.Lett.B335 (1994) 355.

[29] B. Bruegmann and E. Marinari, *More on the exponential bound of four dimensional simplicial quantum gravity* MPI-PhT/94-72, hep-th/9411060.

[30] J. Ambjørn, P. Bialas, Z. Burda and J. Jurkiewicz, Phys.Lett. B342 (1995) 58.

[31] C.Bartocci, U.Bruzzo, M.Carfora and A.Marzuoli, *Entropy of Random Coverings and 4-d Quantum Gravity*, SISSA Ref. 97/94/FM.

[32] B. Durhuus, J. Fröhlich and T. Jonsson, Nucl.Phys. B240 (1984) 453; B257 (1985) 779.

[33] S. Jain and S.D. Mathur, Phys.Lett. B286 (1992) 239.

[34] J. Ambjørn, S. Jain and G. Thorleifsson, Phys.Lett. B307 (1993) 34-39.
Jan Ambjørn, Gudmar Thorleifsson, Phys.Lett.B323 (1994) 7.

[35] A. Nabutovsky and R. Ben-Av, Commun.Math.Phys. 157 (1993) 93.

[36] B.V. de Bakker, *Absence of barriers in dynamical triangulation.* ITFA-94-35, hep-lat - 9411070.

[37] J. Ambjørn, P. Bialas, Z. Burda and J. Jurkiewicz, Phys.Lett. B342 (1995) 58.

[38] C. Holm and W. Janke, Phys.Lett.B335 (194) 143.

[39] W. Bock, J.C. Vink, UCSD-PTH-94-08.

[40] W. Beirl, H. Markum and J. Riedler, Phys.Lett.B341 (1994) 12.

[41] H. Hamber, Nucl.Phys. B400 (1993) 350.

[42] H. Hamber, Phys.Rev. D50 (1994) 3932.

[43] W. Beirl, H. Markum and J. Riedler, Nucl.Phys.B, Proc.Suppl. 34 (1994) 736.

[44] K.S. Stelle, Phys.Rev. D16 (1977) 953.
     E.S. Fradkin and A.A. Tseytlin, Phys.Lett. B104 (1981) 377; B106 (1981) 63;
     Nucl.Phys. B210 (1982) 469.

[45] P.O. Mazur and E. Mottola, Nucl.Phys. B341 (1990) 187.

[46] E. Mottola, *Functional Integration over geometries*, LA-UR-95-80, hep-th-9502109.

[47] E. Witten, Nucl.Phys. B311 (1988/89) 46; B323 (1989) 113.

[48] V.G. Turaev and O.Y. Viro, Topology 31 (1992) 865.

[49] I. Antoniadis and E. Mottola, Phys.Rev. D45 (1992) 2013.

[50] I. Antoniadis, P.O. Mazur and E. Mottola, Nucl.Phys. B388 (1992) 627.

[51] I. Antoniadis, P.O. Mazur and E. Mottola, Phys.Lett. B323 (1994) 284.

[52] I. Antoniadis, private communication.